\documentclass[preprint, showpacs]{revtex4}
\usepackage{graphicx}
\usepackage[bf,SL,BF]{subfigure}
\usepackage{psfrag}
\usepackage{color}
\usepackage{amssymb}
\usepackage{amsmath}
\usepackage{epstopdf}

\newcommand{\be}{\begin{eqnarray}}
\newcommand{\ee}{\end{eqnarray}}

\newcommand{\bfk}{{\bf k}_{\perp}}

\newcommand{\bfp}{{\bf p}_{\perp}}

\newcommand{\bfPhp}{{\bf P}_{h\perp}}
\newcommand{\bfPh}{{\bf P}_{h}}

\newcommand{\avksq}{\langle{k^2}_{\perp}\rangle}

\begin{document}
%%%%%%%%%%%%%%%

\title{%Spin-OAM correlation between Proton and its constituent quark.\\
Sivers and $\cos 2\phi$ Asymmetries in Semi-inclusive Deep Inelastic Scattering  in Light-front Holographic Model}
%in AdS/QCD}
\author{ Tanmay Maji$ ^1$, Dipankar Chakrabarti$ ^1$, Asmita Mukherjee$ ^2$ }
\affiliation{ $ ^1$Department of Physics, 
Indian Institute of Technology Kanpur,
Kanpur 208016, India\\
$ ^2$Department of Physics, 
Indian Institute of Technology Bombay, Powai,
Mumbai 400076, India
}
\date{\today}

\begin{abstract}
The spin asymmetries in SIDIS associated with $T$-odd TMDs are presented in a 
light-front quark-diquark model of a proton. To incorporate the effects of 
the final state interaction, the light front  wave functions are modified to 
have a phase factor which is essential to have Sivers or Boer-Mulders functions.
The Sivers and Boer-Mulder asymmetries are compared with HERMES and COMPASS 
data.
%$\square$ \textbf{At a glance}\\
%$\bullet$ T-odd TMDs: Sivers function and Boer-Mulders functions(considering 
%FSI ).DONE\\
%$\bullet$ Sivers asymmetry and Comparison with HERMES.\\
%$\square$ \textbf{Yet to be done.}\\
%$\bullet$ Comparison of Sivers function with the phenomenological extraction 
%[Anselmino, PRD86, 014028]. DONE\\
%$\bullet$ Scale evolution study for sivers functions.\\
%$\bullet$ Model prediction to Sivers asymmetry and comparison with HERMES 
%data(PRL103,152002). DONE\\
%$\bullet$ Need to edit the section IV. 
%$\bullet$ Boer-Mulders asymmetry and comparison with HERMES. DONE\\
%$\bullet$ Cahn effect/Asymmetry. DONE\\
%$\square$ \textbf{Yet to be done.}\\
%$\bullet$ Error calculation of all the asymmetries.
\end{abstract}
\maketitle
%====================================================
%%%%%%%%%%%%%%%%%%%%%%%%%%%%%%%%%%%%%%
\section{Introduction\label{intro}}
%%%%%%%%%%%%%%%%%%%%%%%%%%%%%%%%%%%%%%
Transverse momentum dependent parton distributions (TMDs) \cite{Anselmino:1994gn, Barone:2001sp} are important to understand the single spin asymmetries (SSA) observed experimentally
 since a long time \cite{Adams:1991rw,Adams:1991cs}. They give a three dimensional picture of the nucleons, together with the generalized  parton distributions (GPDs) and 
represent non-trivial and non-perturbative correlations between the intrinsic transverse momentum of the partons (quarks and gluons) and the spin of the nucleon. 
Of particular interest are the Sivers function \cite{Sivers:1989cc} and the Boer-Mulders function \cite{Boer:1997nt}. At the parton level, Sivers function represents the 
coupling  of the intrinsic transverse momentum of the partons to the transverse spin of the target. It quantifies the 
distribution of unpolarised quarks inside a transversely polarized target.   Sivers effect is particularly interesting as it is sensitive to a phase interference in 
the amplitudes \cite{Brodsky:2002rv} related to the gauge invariance of the underlying QCD interaction \cite{Collins:2002kn,Boer:2003cm,Belitsky:2002sm, Bomhof:2004aw}. Namely, the Sivers 
function is non-zero  only if it takes into account the gluonic initial and 
final state interactions. Such interactions are process dependent. 
The general structure of the process dependence may be quite complicated, however, Sivers function for semi-inclusive deep inelastic scattering (SIDIS)
 is expected to be negative to the Sivers function in Drell-Yan (DY) process
\cite{Belitsky:2002sm,Boer:2003cm}. Sivers effect produces azimuthal asymmetry in SIDIS, that has been measured in experiment by HERMES, 
COMPASS and JLab \cite{Alekseev:2010rw, Alekseev:2008aa,Airapetian:2009ae,Qian:2011py}. Boer-Mulders effect in Drell-Yan process  has been investigated in Fermilab and preliminary results are available \cite{Sbrizzai:2016gro}. There is also recent data from W production at RHIC \cite{Adamczyk:2015gyk}. Another TMD that has gathered considerable interest recently is the Boer-Mulders  function 
\cite{Boer:1997nt}, which gives the distribution of transversely polarized quarks in an unpolarised nucleon. This measures the spin-orbit correlations of quarks. 
Boer-Mulders effect produces a measurable $cos~2 \phi$ azimuthal asymmetry in SIDIS. Like Sivers function, Boer-Mulders function is also process dependent, 
due to the initial and final state interactions.

There have been a lot of phenomenological studies on the Sivers as well as 
Boer-Mulders function (see, for example,
\cite{DAlesio:2004eso,Efremov:2004tp,Anselmino:2005ea,Collins:2005rq,Anselmino:2008sga,Anselmino:2013rya,
Anselmino:2016uie,Martin:2017yms,Barone:2009hw,Zhang:2008nu}).
A lattice calculation is presented in \cite{Musch:2011er}. Extraction of the TMD pdfs from 
experimental data usually relies on the following assumptions
\cite{Anselmino:2016uie}: (i) factorization of the $x$ dependent part of the TMD from the $k_\perp^2$ dependent part , (ii) the $k_\perp$ dependent part is a Gaussian, (iii) in the extraction of the Boer-Mulders function, one usually assumes that it is  proportional to the Sivers function.  The TMD functions are parametrized in terms of several parameters including the average transverse momenta $\langle k_\perp^2 \rangle$ of the partons. This introduces an uncertainty as the experimental values of    
$\langle k_\perp^2 \rangle$ are still not convergent:   
$\langle k_\perp^2 \rangle \approx  0.25~ \mathrm{GeV}^2$ from old EMC data and FNAL SIDIS data, whereas the value is $0.18~ \mathrm{GeV}^2$ derived from HERMES data and this is the value that has been used in the extraction of the Boer-Mulders function. Analysis using a more recent data suggests quite different value; $\langle k_\perp^2 \rangle \approx 0.57 ~ \mathrm{GeV}^2$ (HERMES) and 
$\langle k_\perp^2 \rangle \approx 0.61 ~ \mathrm{GeV}^2$ (COMPASS). A recent extraction
\cite{Martin:2017yms} of the Sivers function, however,  does not use 
any of these values of  $\langle k_\perp^2 \rangle$ as a parameter, but still it uses the factorization between the $x$ dependence and $k_\perp$ dependence. 
The current state of the art can be summarized by saying that the present data is insufficient to confirm the change of sign of the Sivers function between the SIDIS and DY processes, although there is a hint of such sign change from $W^-$ production data at RHIC \cite{Anselmino:2016uie} . 

The Sivers and Boer-Mulders TMDs have also been investigated in various 
phenomenological models \cite{Gamberg:2007wm,Zhang:2008nu, Burkardt:2007xm, 
Pasquini:2010af,Yuan:2003wk,Bacchetta:2003rz,Courtoy:2008vi}. In fact the first 
model calculation of the Sivers asymmetry in \cite{Brodsky:2002rv} showed the 
importance of the phase difference of the overlapping amplitudes to get a 
non-zero asymmetry.  Model studies are also interesting to understand various 
relations between the TMDs and GPDs. An intuitive explanation of the Sivers 
effect was developed in \cite{Burkardt:2003yg} in a model-dependent way. The 
average transverse momentum of an unpolarised quark in a transversely polarized 
nucleon generated due to the Sivers effect 
is related to the distortion in impact parameter space through a lensing 
function, which is the effect of final state interaction.  This relation is 
found to hold in spectator-type models to the lowest non-trivial order, although 
expected to break down when higher order effects are taken into account.   This 
relation is not expected to hold in models where the so-called lensing function 
does not factor out from the GPD in impart parameter space. This relation shows 
the connection of the Sivers function with the orbital angular momentum (OAM) of 
the quarks although depending on the model. A similar model-dependent relation 
is derived in \cite{Meissner:2007rx} between the  Boer-Mulders function, which 
is related to the first derivative of the chiral odd GPDs $\mathcal{E}_T+2 
\tilde H_T$ in the impact parameter space though the lensing function. Sivers 
function and Boer-Mulders function are time-reversal (T) odd  functions whereas 
the GPDs above are T-even, and no model independent relation can be derived 
connecting them. Of course, GPDs and TMDs can be connected through different 
limits of the generalized transverse momentum dependent pdfs (GTMDs). The 
motivation of the present work is to calculate the Sivers and Boer-Mulders 
function using a recently developed quark-diquark model light-front wave 
function of the proton  based on light-front holography, calculate the 
asymmetries to compare with the data  and investigate to what extent the 
model-dependent relations hold.   
      
The light-front quark-diquark model\cite{Maji:2016yqo} has been briefly 
discussed in the next section. The model  has been used to investigate Wigner 
distributions, GPDs  and T-even 
TMDs\cite{Chakrabarti:2017teq,Maji:2017bcz,Maji:2017ill}. The model is also 
found to predict the single-spin asymmetries described by T-even TMDs (Collins 
asymmetries) quite accurately at different experimental scales\cite{Maji:2017zbx}. In this work, the model has been extended to 
incorporate the final state interactions(FSI) into the light front 
wave functions. The FSI generates a phase in the wave function which is 
responsible for non-zero T-odd TMDs i.e., Sivers and Boer-Mulders functions and 
hence the spin asymmetries associated with them. The spin asymmetries evaluated 
in this model are compared with the experimental data using the QCD evolution 
prescribed by Abyat and Rogers\cite{Aybat:2011zv}.

%%%%%%%%%%%%%%%%%%%%%%%%%%%%%%%%%%%%%%%%%%%%%%%%%%%%%%%
\section{light-front quark-diquark model for nucleon\label{model}}
%%%%%%%%%%%%%%%%%%%%%%%%%%%%%%%%%%%%%%%%%%%%%%%%%%%%%%
%In the diquark model, we assume that the virtual incoming photon is interacting with a valence and other two valence quark form a diquark of definite mass with spin-0, called scalar diquark, or with spin-1, called vector diquark. The spin-0 diquarks are in in a flavor singlet state and spin-1 diquarks are in flavor triplet state. 
%The proton state is written
%\be  
%|p; +\rangle = C_s |u~ S^0_0\rangle + C_0^{(u)} |u~ A_0^0\rangle + C_1^{(u)}|u~ A_1^0\rangle + C_0^{(d)} |d~ A_0^1\rangle + C_1^{(d)}|d~ A_1^1\rangle 
%\ee
%as a sum of isoscalar-scalar diquark singlet state $|u~ S^0\rangle$, isoscalar-axial vector diquark state $|u~ A^0\rangle$ and isovector-axial vector diquark $|d~ A^1\rangle$ state\cite{Jakob97,Bacc08}. 
In the light-front quark-diquark model, the proton state is written in a linear 
combination of quark-diquark state with the scalar and axial-vector diquark, 
considering the spin-flavor $SU(4)$ 
structure\cite{Jakob:1997wg,Bacchetta:2008af, Maji:2016yqo} as
\be 
|P; \pm\rangle = C_S|u~ S^0\rangle^\pm + C_V|u~ A^0\rangle^\pm + C_{VV}|d~ A^1\rangle^\pm. \label{PS_state}
\ee
Where, $C_S, C_V$ and $C_{VV}$ are the coefficient of the isoscalar-scalar diquark singlet state $|u~ S^0\rangle$, isoscalar-axial vector diquark state $|u~ A^0\rangle$ and isovector-axial vector diquark state $|d~ A^1\rangle$ respectively. $S$ and $A$ represent the scalar and axial-vector diquark with isospin at their superscript. Under the isospin symmetry, the neutron  state is defined by the above formula with $u\leftrightarrow d$.
%The coefficients $C_i$'s are absorbed in the light front wave function $\psi_{\lambda_q \lambda_D}^{\lambda_N(\nu)}$,Eq.(\ref{LFWF_S}-\ref{LFWF_Vm}) corresponding to the nucleon helicity $\lambda_N$, quark helicity $\lambda_q$ and diquark helicity $\lambda_D$ with flavour index $\nu$.

\begin{figure}
 \includegraphics[width=15cm,clip]{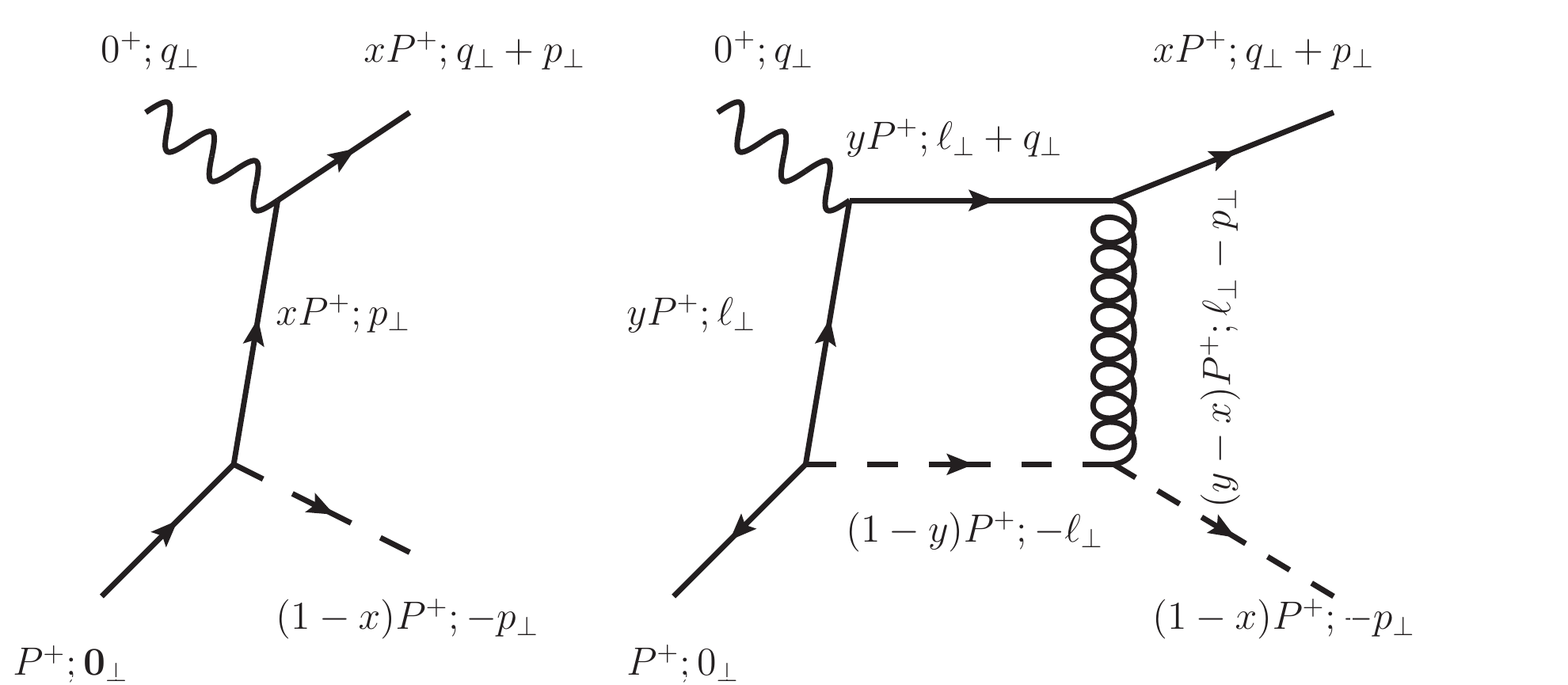}
 \caption{\label{fig_FSI} Left: tree level diagram. Right: FSI diagram for 
$\gamma^* P \to q(qq)$ }
 \end{figure}

The light-cone coordinated $x^\pm=x^0 \pm x^3$. We choose a frame where the incoming proton does not have transverse momentum i,e. $P \equiv \big(P^+,\frac{M^2}{P^+},\textbf{0}_\perp\big)$. However, the struck quark and diquark have equal and opposite transverse momentum:$p\equiv (xP^+, \frac{p^2+|\bfp|^2}{xP^+},\bfp)$ and   $P_X\equiv ((1-x)P^+,P^-_X,-\bfp)$. Here $x=p^+/P^+$ is the longitudinal momentum fraction carried by the struck quark. Detail kinematics of $\gamma^* P \to q(qq)$ are given for tree level and final-state-interaction diagram in the Fig.\ref{fig_FSI}.

The two particle Fock-state expansion for $J^z =\pm1/2$ for spin-0 diquark state is given by
\be
|u~ S\rangle^\pm & =& \int \frac{dx~ d^2\bfp}{2(2\pi)^3\sqrt{x(1-x)}} \bigg[ \psi^{\pm(u)}_{+}(x,\bfp)|+\frac{1}{2}~s; xP^+,\bfp\rangle \nonumber \\
 &+& \psi^{\pm(u)}_{-}(x,\bfp)|-\frac{1}{2}~s; xP^+,\bfp\rangle\bigg],\label{fock_PS}
\ee
where $|\lambda_q~\lambda_S; xP^+,\bfp\rangle$ is the two particle state having struck quark of helicity $\lambda_q$ and a scalar diquark having helicity $\lambda_S=s$(spin-0 singlet diquark helicity is denoted by s to distinguish from triplet diquark). The state with spin-1 diquark is given as \cite{Ellis:2008in}
\be
|\nu~ A \rangle^\pm & =& \int \frac{dx~ d^2\bfp}{2(2\pi)^3\sqrt{x(1-x)}} \bigg[ \psi^{\pm(\nu)}_{++}(x,\bfp)|+\frac{1}{2}~+1; xP^+,\bfp\rangle \nonumber\\
 &+& \psi^{\pm(\nu)}_{-+}(x,\bfp)|-\frac{1}{2}~+1; xP^+,\bfp\rangle +\psi^{\pm(\nu)}_{+0}(x,\bfp)|+\frac{1}{2}~0; xP^+,\bfp\rangle \nonumber \\
 &+& \psi^{\pm(\nu)}_{-0}(x,\bfp)|-\frac{1}{2}~0; xP^+,\bfp\rangle + \psi^{\pm(\nu)}_{+-}(x,\bfp)|+\frac{1}{2}~-1; xP^+,\bfp\rangle \nonumber\\
 &+& \psi^{\pm(\nu)}_{--}(x,\bfp)|-\frac{1}{2}~-1; xP^+,\bfp\rangle  \bigg].\label{fock_PS}
\ee
Where $|\lambda_q~\lambda_D; xP^+,\bfp\rangle$ represents a two-particle state with a quark of helicity $\lambda_q=\pm\frac{1}{2}$ and a axial-vector diquark of helicity $\lambda_D=\pm 1,0(triplet)$.

%%=============================
\section{Final State interaction and T-odd TMDs}
%%===========================
%Brodsky, Hwang and Schmidt(2002) 
The final state interaction\cite{Brodsky:2002cx} produces a non-trivial phase in 
the amplitude and gives non-vanishing T-odd TMDs along with the T-even TMDs. 
There are two T-odd TMDs, $f^{\perp \nu}_{1T}(x,\bfp^2)$ (Sivers function) and $h^{\perp \nu}_1(x,\bfp^2)$ (Boer-Mulders function), at the leading twist. The contribution of FSI is incorporated in the wave functions\cite{Hwang:2010dd} and the wave functions are modified with  spin dependent complex phases as:\\
(i)for scalar diquark
\be 
\psi^{+(u)}_+(x,\bfp)&=& N_S~\bigg[1+i \frac{e_1 e_2}{8 \pi}(\bfp^2 + B)g_1 \bigg] \varphi^{(u)}_{1}(x,\bfp),\nonumber \\
\psi^{+(u)}_-(x,\bfp)&=& N_S\bigg(- \frac{p^1+ip^2}{xM} \bigg) \bigg[1+i \frac{e_1 e_2}{8 \pi}(\bfp^2 + B)g_2\bigg]  \varphi^{(u)}_{2}(x,\bfp),\label{LFWF_S}\\
\psi^{-(u)}_+(x,\bfp)&=& N_S \bigg(\frac{p^1-ip^2}{xM}\bigg) \bigg[1+i \frac{e_1 e_2}{8 \pi}(\bfp^2 + B)g_2\bigg] \varphi^{(u)}_{2}(x,\bfp),\nonumber \\
\psi^{-(u)}_-(x,\bfp)&=&  N_S~ \bigg[1+i \frac{e_1 e_2}{8 \pi}(\bfp^2 + B)g_1\bigg]\varphi^{(u)}_{1}(x,\bfp),\nonumber
\ee
 (ii) for axial-vector diquark(for $J=\pm 1/2$ ) 
\be 
\psi^{+(\nu)}_{+~+}(x,\bfp)&=& N^{(\nu)}_1 \sqrt{\frac{2}{3}} \bigg(\frac{p^1-ip^2}{xM}\bigg)\bigg[1+i \frac{e_1 e_2}{8 \pi}(\bfp^2 + B)g_2 \bigg] \varphi^{(\nu)}_{2}(x,\bfp),\nonumber \\
\psi^{+(\nu)}_{-~+}(x,\bfp)&=& N^{(\nu)}_1 \sqrt{\frac{2}{3}} \bigg[1+i \frac{e_1 e_2}{8 \pi}(\bfp^2 + B)g_1 \bigg] \varphi^{(\nu)}_{1}(x,\bfp),\nonumber \\
\psi^{+(\nu)}_{+~0}(x,\bfp)&=& - N^{(\nu)}_0 \sqrt{\frac{1}{3}} \bigg[1+i \frac{e_1 e_2}{8 \pi}(\bfp^2 + B)g_1 \bigg] \varphi^{(\nu)}_{1}(x,\bfp),\label{LFWF_Vp}\\
\psi^{+(\nu)}_{-~0}(x,\bfp)&=& N^{(\nu)}_0 \sqrt{\frac{1}{3}} \bigg(\frac{p^1+ip^2}{xM} \bigg) \bigg[1+i \frac{e_1 e_2}{8 \pi}(\bfp^2 + B)g_2 \bigg] \varphi^{(\nu)}_{2}(x,\bfp),\nonumber \\
\psi^{+(\nu)}_{+~-}(x,\bfp)&=& 0,\nonumber \\
\psi^{+(\nu)}_{-~-}(x,\bfp)&=&  0, \nonumber 
\ee
and for $J=-1/2$
\be 
\psi^{-(\nu)}_{+~+}(x,\bfp)&=& 0,\nonumber \\
\psi^{-(\nu)}_{-~+}(x,\bfp)&=& 0,\nonumber \\
\psi^{-(\nu)}_{+~0}(x,\bfp)&=& N^{(\nu)}_0 \sqrt{\frac{1}{3}} \bigg( \frac{p^1-ip^2}{xM} \bigg) \bigg[1+i \frac{e_1 e_2}{8 \pi}(\bfp^2 + B)g_2 \bigg] \varphi^{(\nu)}_{2}(x,\bfp),\label{LFWF_Vm}\\
\psi^{-(\nu)}_{-~0}(x,\bfp)&=& N^{(\nu)}_0\sqrt{\frac{1}{3}} \bigg[1+i \frac{e_1 e_2}{8 \pi}(\bfp^2 + B)g_1 \bigg] \varphi^{(\nu)}_{1}(x,\bfp),\nonumber \\
\psi^{-(\nu)}_{+~-}(x,\bfp)&=& - N^{(\nu)}_1 \sqrt{\frac{2}{3}} \bigg[1+i \frac{e_1 e_2}{8 \pi}(\bfp^2 + B)g_1 \bigg] \varphi^{(\nu)}_{1}(x,\bfp),\nonumber \\
\psi^{-(\nu)}_{-~-}(x,\bfp)&=& N^{(\nu)}_1 \sqrt{\frac{2}{3}} \bigg(\frac{p^1+ip^2}{xM}\bigg) \bigg[1+i \frac{e_1 e_2}{8 \pi}(\bfp^2 + B)g_2 \bigg] \varphi^{(\nu)}_{2}(x,\bfp),\nonumber
\ee
Where,
\be 
g_1 &=& \int^1_0 d\alpha \frac{-1}{\alpha(1-\alpha)\bfp^2 + \alpha m_g^2 + (1-\alpha)B},\\
g_2 &=& \int^1_0 d\alpha \frac{-\alpha}{\alpha(1-\alpha)\bfp^2 + \alpha m_g^2 + (1-\alpha)B}\\
{\rm and},\nonumber\\
B &=& x(1-x)(-M^2+\frac{m^2_q}{x}+\frac{m^2_D}{1-x}).
\ee
$M, m_q,m_D$ and $ m_g$ are mass of proton, struck quark, diquark and gluon respectively. We take $m_g=0$ at the end of the calculations. $N_S, N^{(\nu)}_0$ and $N^{(\nu)}_1$  are the normalization constants. 

The LFWFs $\varphi^{(\nu)}_i(x,\bfp)$ are modified form of the  soft-wall AdS/QCD prediction as\cite{Gutsche:2013zia}
\be
\varphi_i^{(\nu)}(x,\bfp)=\frac{4\pi}{\kappa}\sqrt{\frac{\log(1/x)}{1-x}}x^{a_i^\nu}(1-x)^{b_i^\nu}\exp\bigg[-\delta^\nu\frac{\bfp^2}{2\kappa^2}\frac{\log(1/x)}{(1-x)^2}\bigg],
\label{LFWF_phi}
\ee
introducing the parameters $a^{\nu}_i, b^\mu_i$ and $\delta^\nu$. The wave 
functions $\varphi_i^\nu ~(i=1,2)$ reduce to the AdS/QCD 
prediction\cite{Brodsky:2007hb} for the parameters $a_i^\nu=b_i^\nu=0$  and 
$\delta^\nu=1.0$.
 We use the AdS/QCD scale parameter $\kappa =0.4~GeV$ as determined in \cite{Chakrabarti:2013gra} and the quarks are  assumed  to be  massless. 

The unintegrated quark-quark correlator for polarized SIDIS is defined  as 
%\begin{align}
\be
\Phi^{\nu [\Gamma]}(x,\textbf{p}_{\perp};S)&=&\frac{1}{2}\int \frac{dz^- d^2z_T}{2(2\pi)^3} e^{ip.z} \langle P; S|\overline{\psi}^\nu (0)\Gamma \mathcal{W}_{[0,z]} \psi^\nu (z) |P;S\rangle, \label{TMD_cor}
\ee
%\end{align} 
with a flavour $\nu$. The summations over the color indices of quarks are implied. $ \mathcal{W}_[0,z]$ is the Wilson line, the effect of which is incorporated in the
LFWFs in terms of FSI. Here, $p$ is the momentum of the struck quark inside the nucleon of momentum P, spin S and $x ~(x=p^+/P^+)$ is the longitudinal momentum fraction carried by struck quark. 
%We choose a frame where the nucleon is collinear with photon and give a  transverse momentum to the final state hadron $P_h$. 
% and a frame where the nucleon momentum and quark momentum are $ P\equiv (P^+,\frac{M^2}{P^+},\textbf{0} ),~ q\equiv (x_B P^+, \frac{Q^2}{x_BP^+},\textbf{0})$ respectively, $x_B= \frac{Q^2}{2P.q}$ is the Bjorken scaling with $Q^2 = -q^2$. 
We choose the light-cone gauge $A^+=0$.
The nucleon with helicity $\lambda_N$ has  spin components $S^+ = \lambda_N \frac{P^+}{M},~ S^- = \lambda_N \frac{P^-}{M},$ and $ S_T $.  
At leading twist, the T-odd TMDs are defined as
\begin{eqnarray}
\Phi^{\nu [\gamma^+]}(x,\textbf{p}_{\perp};S)&=& ... - \frac{\epsilon^{ij}_Tp^i_\perp S^j_T}{M}f^{\perp  \nu} _{1T}(x,\textbf{p}_{\perp}^2),\label{Phi_1}\\
%\Phi^{\nu [\gamma^+ \gamma^5]}(x,\textbf{p}_{\perp};S) &=&  \lambda g_{1L}^\nu (x,\textbf{p}_{\perp}^2) + \frac{\textbf{p}_{\perp}.\textbf{S}_T}{M} g^\nu _{1T}(x,\textbf{p}_{\perp}^2),\label{Phi_2}\\
\Phi^{\nu [i \sigma^{j +}\gamma^5]}(x,\textbf{p}_{\perp};S)& = & 
%S^j_T h_1^\nu (x,\textbf{p}_{\perp}^2) + \lambda\frac{p^j_\perp}{M}h^{\perp  \nu} _{1L}(x,\textbf{p}_{\perp}^2)\nonumber\\
%&&+ \frac{2 p^j_\perp \textbf{p}_{\perp}.\textbf{S}_T - S^j_T \textbf{p}^2_{\perp}}{2M^2} h^{\perp  \nu} _{1T}(x,\textbf{p}_{\perp}^2)
... + \frac{\epsilon_T^{ij}p^i_{\perp}}{M}h^{\perp  \nu} 
_1(x,\textbf{p}_{\perp}^2),\label{Phi_3}
\end{eqnarray}
where the ellipses indicate the terms involving  T-even TMDs.

Using the Eq.(\ref{PS_state}) in the Eq.(\ref{TMD_cor}) the correlators for 
transversely polarized proton are written in terms of overlap representations as
\be 
%-\frac{p^2_\perp}{M} f^\perp_{1T}(x,\bfp)
\Phi^{\nu [\gamma^+]}(x,\textbf{p}_{\perp};\uparrow)&=& \frac{1}{2} \bigg[ C^2_S \frac{1}{16 \pi^3} \sum_{\lambda_q}\sum_{\lambda_N} \sum_{\lambda^\prime_N} \psi^{\lambda_N \dagger}_{\lambda_q}(x,\bfp)\psi^{\lambda^\prime_N}_{\lambda_q}(x,\bfp)\bigg]^\nu \nonumber\\
&&\hspace{1.5cm} + \frac{1}{2} \bigg[ C^2_A \frac{1}{16 \pi^3} \sum_{\lambda_q} \sum_{\lambda_D}\sum_{\lambda_N} \sum_{\lambda^\prime_N} \psi^{\lambda_N \dagger}_{\lambda_q \lambda_D}(x,\bfp)\psi^{\lambda^\prime_N}_{\lambda_q \lambda_D}(x,\bfp)\bigg]^\nu, \label{OLR_Siv}\\
\Phi^{\nu [i \sigma^{1 +}\gamma^5]}(x,\textbf{p}_{\perp};\uparrow)&=& \frac{1}{2} \bigg[ C^2_S \frac{1}{16 \pi^3} \sum_{\lambda_q}\sum_{\lambda^\prime_q} \sum_{\lambda_N} \psi^{\lambda_N \dagger}_{\lambda_q}(x,\bfp)\psi^{\lambda_N}_{\lambda^\prime_q}(x,\bfp)\bigg]^\nu \nonumber\\
&&\hspace{1.5cm} + \frac{1}{2} \bigg[ C^2_A \frac{1}{16 \pi^3} \sum_{\lambda_q}\sum_{\lambda^\prime_q} \sum_{\lambda_D}\sum_{\lambda_N}  \psi^{\lambda_N \dagger}_{\lambda_q \lambda_D}(x,\bfp)\psi^{\lambda_N}_{\lambda^\prime \lambda_D}(x,\bfp)\bigg]^\nu. \label{OLR_BM}
\ee 
Where, $\lambda_q,\lambda_D=\pm$ represent the helicity of quark and diquark respectively. In the Eq.\ref{BM_TMD} the quark polarization is taken along $x$-axis, $j=1$.  The first term in the right-hand-side is for the scalar diquark and the second term is corresponding to the vector diquark.
Note, the first terms in the right-hand-side of the two Eqs.(\ref{OLR_Siv},\ref{OLR_BM}) become zero for d quark as $N^d_S=0$ in the scalar wave functions. $C^2_A$ in the second term stands for the coefficients $C^2_V$ and $C^2_{VV}$ for u quark and d quark respectively.
Comparing Eqs.(\ref{Phi_1},\ref{Phi_3}) with the Eqs.(\ref{OLR_Siv},\ref{OLR_BM}) the Sivers function $f_{1T}^{\perp\nu}(x,\bfp^2)$ and Boer-Mulders functions can be written in the LFQDM as
\be
f_{1T}^{\perp \nu}(x,\bfp^2)&=& \bigg(C^2_S N^{\nu 2}_S -C^2_A \frac{1}{3}N^{\nu 2}_0 \bigg) f^\nu(x,\bfp^2), \label{siv_TMD}\\
h_{1}^{\perp \nu}(x,\bfp^2)&=& \bigg(C^2_S N^{\nu 2}_S + C^2_A \big(\frac{1}{3}N^{\nu 2}_0 + \frac{2}{3}N^{\nu 2}_1\big)\bigg) f^\nu(x,\bfp^2). \label{BM_TMD}
\ee
Where
\be 
f^\nu(x,\bfp^2) &=& - C_F \alpha_s \bigg[\bfp^2 + x(1-x)(-M^2+\frac{m_D^2}{1-x}+\frac{m_q^2}{x})\bigg] \frac{1}{\bfp^2} \nonumber \\ 
&& \hspace{1cm} \times \ln\bigg[1+\frac{\bfp^2}{ x(1-x)(-M^2+\frac{m_D^2}{1-x}+\frac{m_q^2}{x})}\bigg] \nonumber\\
&& \hspace{2cm} \times \frac{\ln(1/x)}{\pi \kappa^2}
 x^{a^{\nu}_1 +a^{\nu}_2-1}(1-x)^{b^{\nu}_1+b^{\nu}_2-1} \exp\bigg[- \delta^\nu \frac{\bfp^2\ln(1/x)}{\kappa^2(1-x)^2}\bigg],
\ee
with struck quark mass $m_q$ and diquark mass $m_D$. In the final state
interaction, gluon exchange strength $\frac{e_1e_2}{4\pi} \rightarrow - 
C_F\alpha_s$. Here the $e_1$ and $e_2$ are color charge of the struck quark and diquark respectively. The values of the parameters $a^\nu_i, b^\nu_i~(i=1,2)$ and $\delta^\nu$ are given in \cite{Maji:2017bcz}.

\begin{figure}[htbp]
\begin{minipage}[c]{0.98\textwidth}
\includegraphics[width=7.5cm,clip]{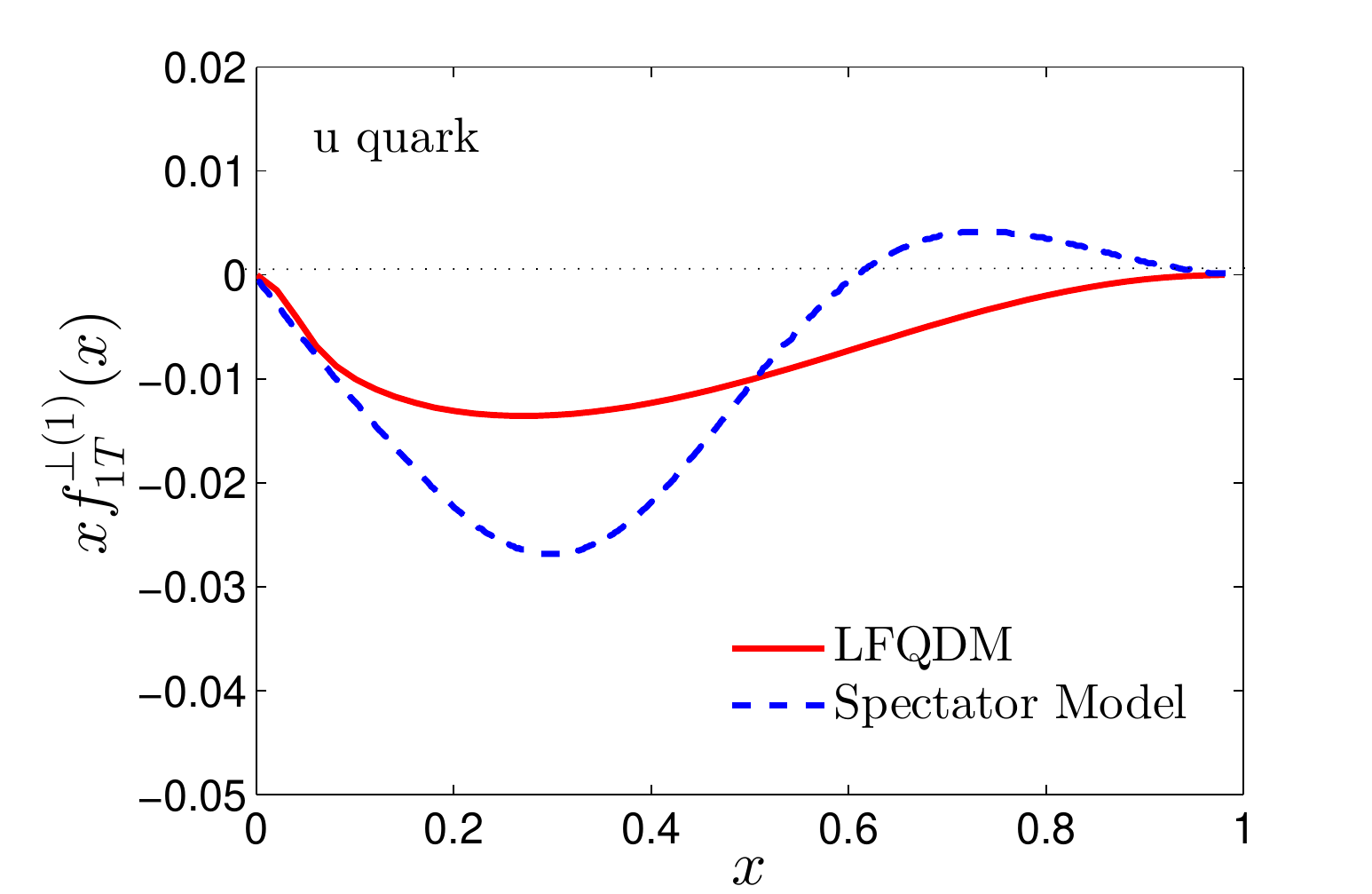}
\includegraphics[width=7.5cm,clip]{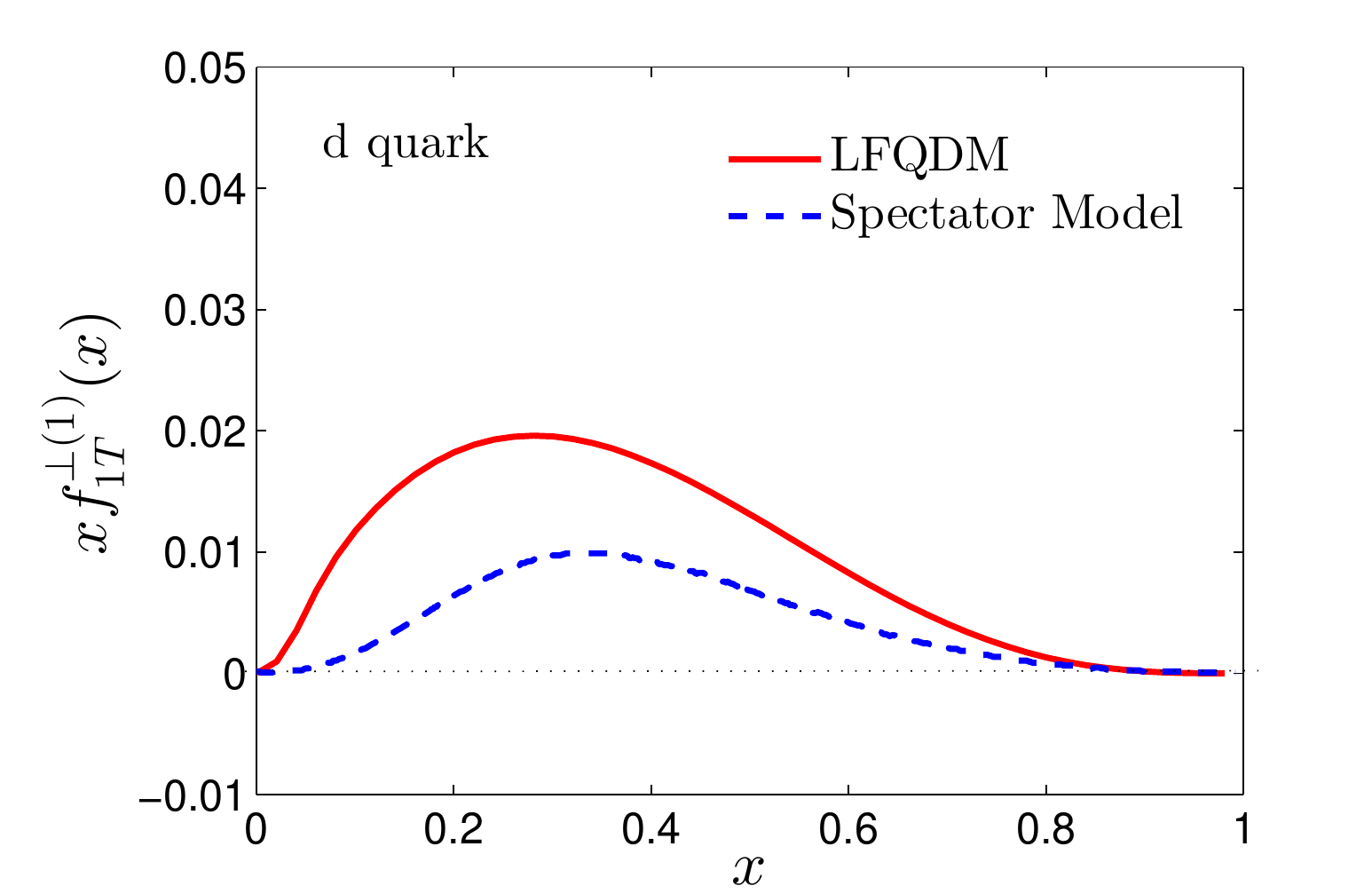}
\end{minipage}
\caption{\label{fig_siv_mu0} $xf^{\perp(1)}_{1T}(x)$ are for $u$   and 
 $d$ quarks %with a regulator $m^2_q=0.003~ GeV^2$ and  $m^2_D=0.8~GeV^2$
at initial scale $\mu_0=0.8 ~GeV$. Red continuous lines represent the model result in LFQDM and blue dashed line represent the result in spectator model \cite{Bacchetta:2008af}.}
\end{figure} 

\begin{figure}[htbp]
\begin{minipage}[c]{0.98\textwidth}
\includegraphics[width=7.5cm,clip]{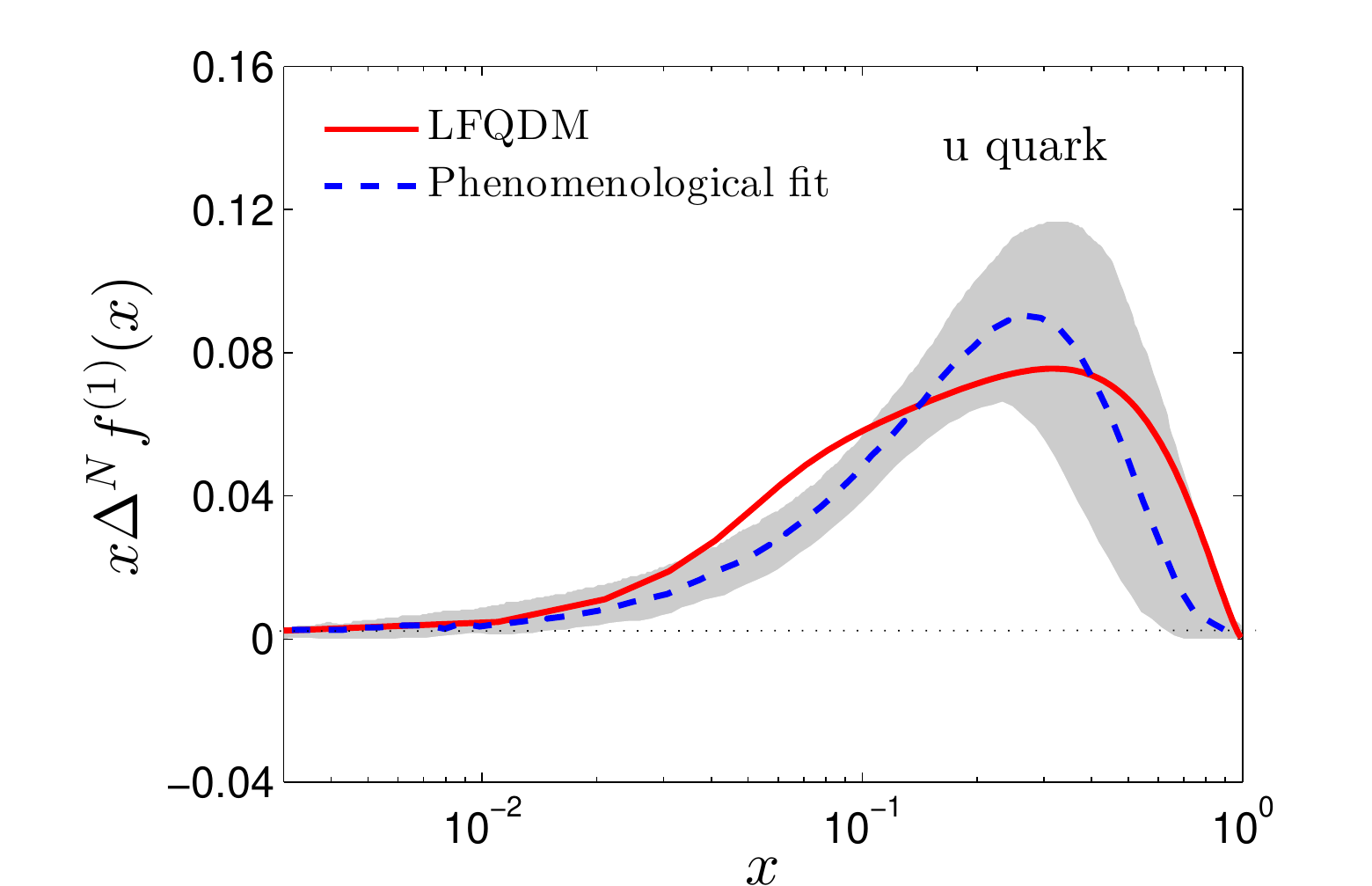}
\includegraphics[width=7.5cm,clip]{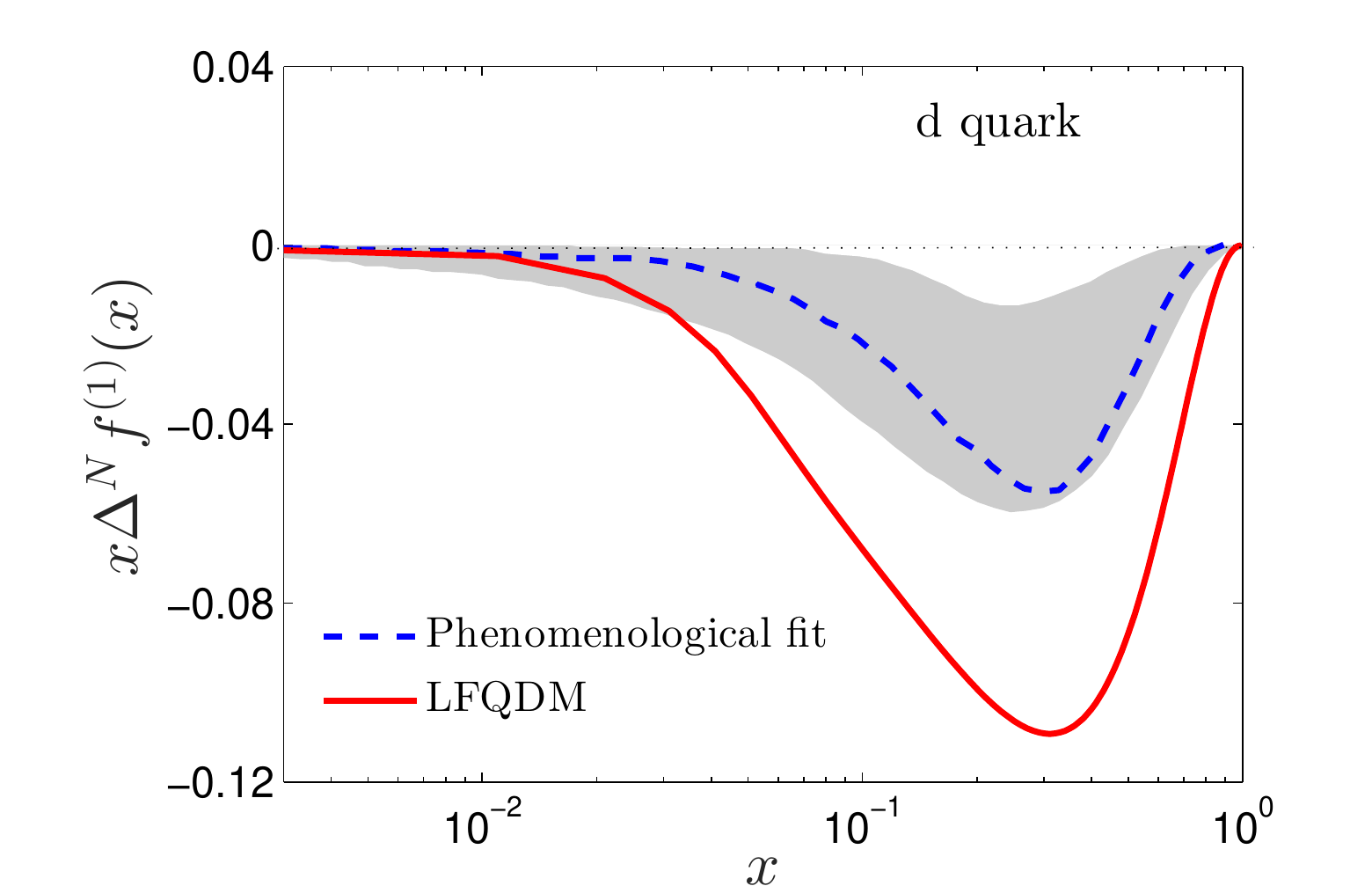} 
\end{minipage}
\caption{\label{fig_siv_1GeV} $x\Delta^N f^{(1)}(x)$ are shown for u and d quarks %with a regulator $m^2_q=0.003~ GeV^2$ and  $m^2_D=0.8~GeV^2$ 
at the scale $\mu=1 ~GeV$. Our model result is shown in red continuous lines. Blue dashed lines represent the phenomenological extraction \cite{Anselmino:2012aa} from the best fit of the Sivers asymmetries measured by HERMES \cite{Airapetian:2009ae} and COMPASS \cite{Anselmino:2011gs, Alekseev:2008aa} collaborations.}
\end{figure} 

Moment of the Sivers functions, defined as 
\be
f^{\perp(1)}_{1T}(x) = \int d^2 p_\perp \frac{p^2_\perp}{2 M^2} f^{\perp}_{1T}(x,\bfp^2),
%h^{\perp(1)}_{1}(x) = \int d^2 p_\perp \frac{p^2_\perp}{2 M^2} h^{\perp}_{1}(x,\bfp^2),  
\ee
are shown in Fig.\ref{fig_siv_mu0} at the initial scale and compared with the spectator model\cite{Bacchetta:2008af}. Our model result for u quark does not have any positive peak like the spectator model. The scale evolution of the distributions are not included in the spectator model.

 According to Burkardt sum rule\cite{Burkardt:2004ur}, the net transverse Sivers 
 momentum when summed over all the constituents is zero. In the 
quark-diquark model, the constituents are quarks($q$) and diquarks($D$) only 
and the statement can be written as:
 \be
 \sum_{i=q,D}\langle k_\perp^i\rangle=0.
 \ee
  The sum rule in terms of Sivers function can be written 
as\cite{Efremov:2004tp}
  \be
  \sum_{i=q,D}\int dxf^{\perp(1)}_{1T}(x) =0.
  \ee
In a scalar diquark model, it was shown\cite{Goeke:2006ef} that the Sivers functions
for the quark and diquark are related by
\be f_{1T}^{\perp D}(x,\bfp^2)=-f_{1T}^{\perp q}(1-x,\bfp^2). \ee
The same relation also holds in our model when averaged over the vector diquark 
polarizations and the Burkardt sum rule is satisfied.

In Fig.\ref{fig_siv_1GeV}, the $x\Delta^N f^{(1)}(x)$ are presented at the scale $\mu=1~GeV$ and compared with the phenomenological fit from the HERMES and COMPASS data. The moment of the Sivers function $\Delta^N f^{(1)}(x)$ is defined as
\be 
\Delta^N f^{(1)}(x)= \int d^2p_\perp (\frac{p_\perp}{4 M}) \Delta^N f_{\nu/P^\uparrow}(x,\bfp),
\ee 
where
\be 
\Delta^N f_{\nu/P^\uparrow}(x,\bfp)=(-\frac{2 p_\perp}{ M}) f^\perp_{1T}(x,\bfp^2).
\ee

\begin{figure}[htbp]
\begin{minipage}[c]{0.98\textwidth}
\includegraphics[width=7.5cm,clip]{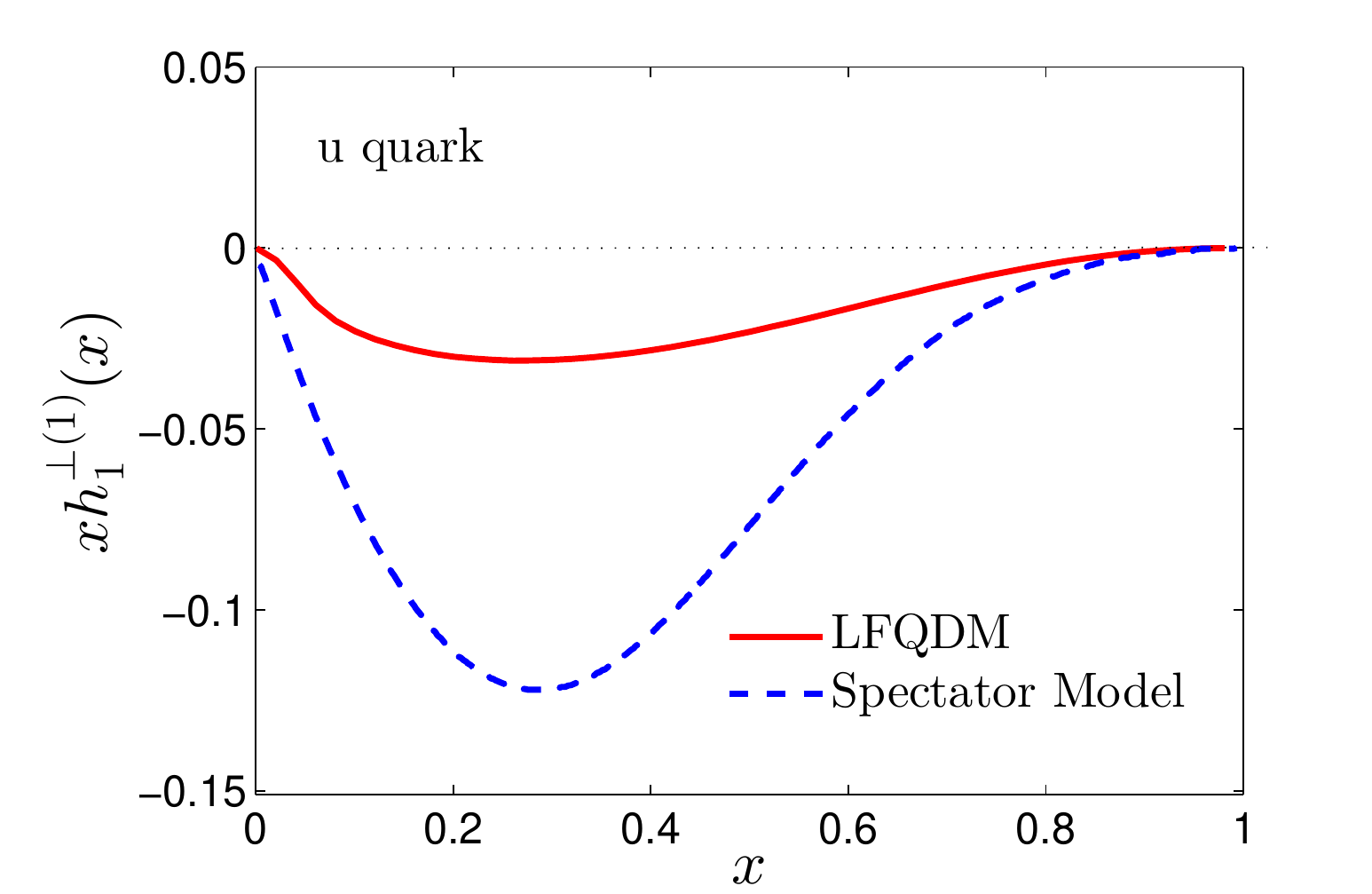}
\includegraphics[width=7.5cm,clip]{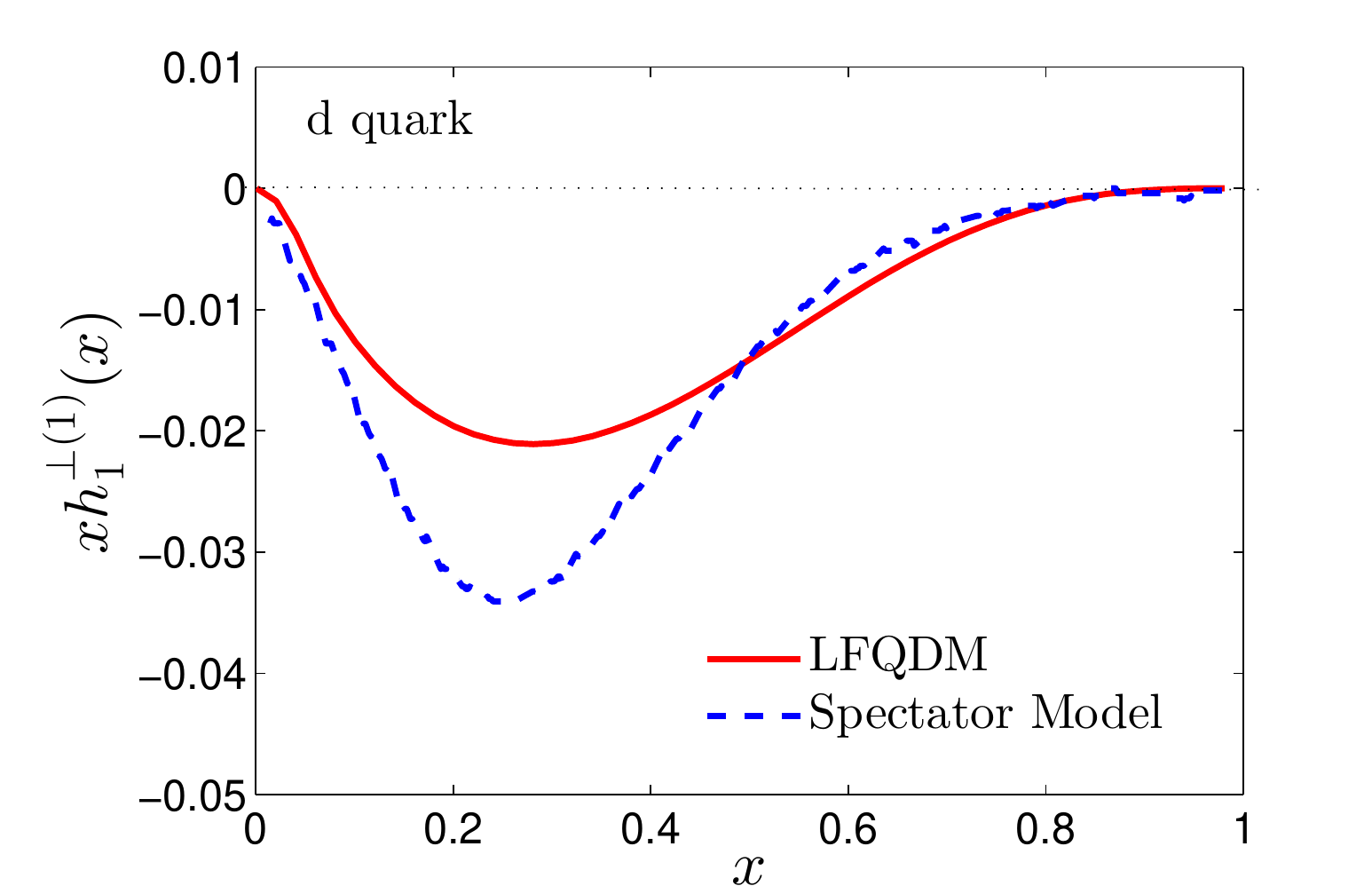}
\end{minipage}
\caption{\label{fig_BM_mu0} $xh^{\perp(1)}_{1}(x)$ are shown for u and d quarks %with a regulator $m^2_q=0.003~ GeV^2$ and  $m^2_D=0.8~GeV^2$
at initial scale $\mu_0=0.8 ~GeV$. Red continuous lines represent the model result in LFQDM and blue dashed line represent the result in spectator model \cite{Bacchetta:2008af}.}
\end{figure} 

In Fig.\ref{fig_BM_mu0}, we show our model result for moment of Boer-Mulder functions, defined as
\be
h^{\perp(1)}_{1}(x) = \int d^2 p_\perp \frac{p^2_\perp}{2 M^2} h^{\perp}_{1}(x,\bfp^2),  
\ee
at the initial scale and compare with the spectator model. 

In this model, we observe  
\be 
|h^\perp_1(x,\bfp^2)|> |f^\perp_{1T}(x,\bfp^2)|.
\ee
From Eq.(\ref{siv_TMD}) and Eq.(\ref{BM_TMD}), we can easily see that 
Boer-Mulders function is proportional to the Sivers function. In fact,
Boer-Mulders function is parametrized\cite{Barone:2009hw} as 
\be 
h^{\perp\nu}_1(x,\bfp^2) \simeq \lambda^\nu f^{\perp\nu}_{1T}(x,\bfp^2).\label{Lq}
\ee 
The Table  \ref{tab_lam} shows our model result of $\lambda^\nu$ and compared 
with the result of HERMES and COMPASS data fits \cite{Barone:2009hw} for $\cos 
2\phi$ asymmetry in SIDIS. The results indicate that Boer-Mulders functions are 
negative for both u and d quarks.

\begin{table}[ht]
\centering % used for centering table 
\begin{tabular}{|c|c|c|c|c|c|}
 \hline
 ~~  ~~&~~  $\lambda^u$ ~~&~~ $\lambda^d$\\ \hline
 ~~ LFQDM ~~&~~ $ 2.29 $ ~~&~~ $-1.08$ \\ 
 ~~ Phenomenological fit  ~~&~~ $2.1 \pm0.1$ ~~&~~ $-1.11\pm0.02$\\ \hline
 \end{tabular} 
\caption{$\lambda^\nu$ of Eq.(\ref{Lq}) for u and d quarks are shown in our model and fitted data \cite{Barone:2009hw} of HERMES and COMPASS.} % title of Table 
\label{tab_lam} % is used to refer this table in the text 
\end{table}

%The first moment of Sivers and Boer-Mulders functions defined as
%\be
%f^{\perp(1)}_{1T}(x) = \int d^2 p_\perp \frac{p^2_\perp}{2 M^2} f^{\perp}_{1T}(x,\bfp^2),\\
%h^{\perp(1)}_{1}(x) = \int d^2 p_\perp \frac{p^2_\perp}{2 M^2} h^{\perp}_{1}(x,\bfp^2),  
%\ee
% are shown in Fig.\ref{fig_siv_BM}.

%%%==================================
\section{Sivers asymmetry and Boer-Mulders asymmetry}
%%%%+==============================
The Sivers Asymmetry correlates between transverse momentum of parton and transverse polarization of nucleon.  In the SIDIS precesses, Sivers asymmetry can be extracted by incorporating the weight factor $\sin(\phi_h-\phi_S)$ as 
\be 
A^{\sin(\phi_h-\phi_S)}_{UT}&=&\frac{\int d\phi_h d\phi_S [d\sigma^{\ell P^\uparrow \to \ell' h X}-d\sigma^{\ell P^\downarrow \to \ell' h X}]\sin(\phi_h-\phi_S)}{\int d\phi_h d\phi_S [d\sigma^{\ell P^\uparrow \to \ell' h X}+d\sigma^{\ell P^\downarrow \to \ell' h X}]} \label{Asy_def}\\
\ee
Where $\uparrow,\downarrow$ at the superscript of $P$ represent the up and down transverse spin of the target proton. According to the QCD factorization scheme the Semi-Inclusive Deep Inelastic Scattering(SIDIS) cross-section for the one photon exchange process
$\ell N \to \ell' h X$ is written as
\be 
d\sigma^{\ell N \to \ell' h X}=\sum_\nu \hat{f}_{\nu/P}(x,\bfp;Q^2)\otimes d\hat{\sigma}^{\ell q \to \ell q} \otimes \hat{D}_{h/\nu}(z,\bfk;Q^2).
\ee  
Where the second term represents the 
hard scattering part which is calculable in pQCD.  The soft part is factorized 
into TMDs, denoted by $\hat{f}_{\nu/P}(x,\bfp;Q^2)$ and fragmentation functions 
(FF), denoted by $\hat{D}_{h/\nu}(z,\bfk;Q^2)$.  This scheme holds in small 
$\bfPhp$ and large $Q$ region, 
$P_{h\perp}^2 \simeq \Lambda^2_{QCD} \ll Q^2 $. The quark-gluon corrections and higher order pQCD corrections become important at large $\bfPhp$ regime \cite{Bacchetta:2008af, Ji:2006br, Anselmino:2006rv}. 
%The TMD factorization theorem is not proven generically for all the process. However, a 
The TMD factorization is presented for the 
SIDIS and the DY processes in \cite{Ji:2004wu,Ji:2004xq, 
Collins,GarciaEchevarria:2011rb,Echevarria:2012js} and latter on used in 
\cite{Aybat:2011zv,Aybat:2011ge,Aybat:2011ta,Anselmino:2012aa}.
%The kinematics of SIDIS are given in Fig.\ref{frame}.
The kinematic variables are defined in  the $\gamma^*-N$ center of mass frame as 
\be 
x=\frac{Q^2}{2(P.q)}=x_B, \hspace{1.5cm}
z=\frac{P.P_h}{P.q}=z_h, \hspace{1.5cm}
y=\frac{P.q}{P.\ell}=\frac{Q^2}{s x} .
\ee 
%In this frame, struck quark and diquark have equal and opposite transverse momentum and produced hadron gets a non-zero transverse momentum. 
 Bjorken scaling $x_B=\frac{Q^2}{2P.q}$ with $Q^2=-q^2$. The fractional energy 
transferred by the photon in the lab system is $y$ and the
energy fraction carried by the produced hadron is $z=\bfPh^-/k^-$. The transverse momentum of the fragmenting quark is denoted as $\bfk$.  The momentum of the virtual photon $q \equiv (x_B P^+,\frac{Q^2}{x_B P^+}, \textbf{0}_\perp)$  and of the incoming proton $P \equiv (P^+, \frac{M^2}{P^+}, \textbf{0}_\perp)$. 
The struck quark have non-zero transverse  momentum $\bfp$ with the momentum $p \equiv (xP^+, \frac{p^2+|\bfp|^2}{xP^+}, \bfp)$ and the diquark carries $p_D \equiv ((1-x)P^+, \frac{p^2+|\bfp|^2}{(1-x)P^+}, -\bfp)$. In this frame, the produced hadron has a finite transverse momentum $\bfPhp$.
% and the components are $\bfPh \equiv (P^+, P^-, \bfPhp)$. 
%We use the light-cone convention $x^\pm = x^0 \pm x^3 $. 
%In this frame, though the incoming proton dose not have transverse momentum, the constituent quarks can have  non-zero transverse momenta which sum up to zero. $\bfp, \bfk$ and $\bfPhp$ are the transverse momentum  carried by struck quark, fragmenting quark and fragmented hadron respectively.
 At $\mathcal{O}(\bfp/Q)$, the relation between $\bfp, \bfk$ and $\bfPhp$ is given as  
$\bfk=\bfPhp-z\bfp$.
%Here we consider one photon interaction only. 
The transverse momentum of produced hadron makes an azimuthal angle $\phi_h$ with respect to the 
lepton plane and transverse spin($S_P$) of the proton has an azimuthal angle $\phi_S$.
%
%\begin{figure}[htbp]
%\includegraphics[width=7.2cm,clip]{frame.jpg}
%\caption{\label{frame} $\gamma^*-P$ center of mass frame: produced hadron has a non-zero transverse momentum($\bfPhp$) in this frame and makes an azimuthal angle of $\phi_h$. The proton spin ($S$) has an azimuthal angle of $\phi_S$. All kinematics are given in text.}
%\end{figure}
Then the SIDIS cross-section deference \cite{Anselmino:2011ch} in the numerator can be written as  
\be 
\frac{d\sigma^{\ell P^\uparrow \to \ell' h X}-d\sigma^{\ell P^\downarrow \to \ell' h X}}{dx_B dy dz d^2\bfPhp d\phi_S}&=& \frac{2\alpha^2}
{s x y^2}2\bigg[\frac{1+(1-y)^2}{2}\sin(\phi_h-\phi_S)F^{\sin(\phi_h-\phi_S)}_{UT}\nonumber\\
+&&\!\!\!\!\!\!\!\! (1-y)\bigg(\sin(\phi_h+\phi_S)F^{\sin(\phi_h+\phi_S)}_{UT}
+\sin(3\phi_h-\phi_S)F^{\sin(3\phi_h-\phi_S)}_{UT}\bigg)\nonumber\\
+&&\!\!\!\!\!\!\! (2-y)\sqrt{(1-y)}\bigg(\sin\phi_S 
F^{\sin\phi_S}_{UT}+\sin(2\phi_h-\phi_S)F^{\sin(2\phi_h-\phi_S)}_{UT}\bigg)\bigg
].\label{N_UT}
\ee
The weighted structure functions, $F^{\mathcal{W}(\phi_h,\phi_S)}_{S_\ell 
S}$, are defined as 
\be 
F^{\mathcal{W}(\phi_h,\phi_S)}_{S_\ell S}%&=&\mathcal{C}[\mathcal{W} 
%\hat{f}(x,\bfp) \hat{D}(z,\bfk)] \nonumber\\
= \sum_\nu e^2_\nu  \int d^2\bfp d^2\bfk \delta^{(2)}(\bfPhp-z\bfp-\bfk) 
\mathcal{W}(\bfp,\bfPhp) \hat{f}^\nu(x,\bfp)\hat{D}^\nu(z,\bfk),\label{conv}
\ee
where  $\hat{f}^\nu(x,\bfp)$ and $\hat{D}^\nu(z,\bfk)$ represent leading twist 
TMDs and FFs respectively.
 Integrating  the numerator over $\phi_h$ and $\phi_S$, with a particular weight factor $\mathcal{W}(\phi_h,\phi_S)$, one can project out the corresponding structure function $F^{\mathcal{W}(\phi_h,\phi_S)}_{S_\ell S}$ and hence the particular asymmetry can be found. For example, the $\phi_h$ and $\phi_S$ integration with the weight factors $\sin(\phi_h-\phi_S)$, and $ \cos(2\phi_h)$  end up with the Sivers asymmetry and Boer-Mulders asymmetry. 
%The second term  corresponds to the Collins asymmetry which has contribution from transversity TMD ($h^\nu_1$) and Collins fragmentation 
%function($H^{\perp h/\nu}_{1}$). The third term has contribution from pretzelocity distribution($h^{\perp\nu}_{1T}$). 
%The fourth and fifth terms have contributions from multiple TMDs and FFs. Among these five SSAs, only two of them
%$A^{\sin(\phi_h+\phi_S)}_{UT}(x,z,\bfPhp,y)$ and $A^{\sin(3\phi_h-\phi_S)}_{UT}(x,z,\bfPhp,y)$,  involve T-even TMDs and will be discussed here.

Similarly, the denominator can be written as 
\be 
\frac{d\sigma^{\ell P^\uparrow \to \ell' h X} + d\sigma^{\ell P^\downarrow \to \ell' h X}}{dx_B dy dz d^2\bfPhp d\phi_S}&=& \frac{2\alpha^2}{s x y^2}2 \bigg[\frac{1+(1-y)^2}{2}F_{UU}+(2-y)\sqrt{1-y}\cos\phi_h F^{\cos\phi_h}_{UU} \nonumber\\
&+&(1-y)cos2\phi_h F^{\cos2\phi_h}_{UU}\bigg]. \label{D_UT}
\ee
Thus Sivers asymmetry can be written in terms of
structure functions \cite{Anselmino:2011ch} as
\be 
A^{\sin(\phi_h-\phi_S)}_{UT}(x,z,\bfPhp,y)&&= 
\frac{2\pi^2\alpha^2\frac{1+(1-y)^2}{s x y^2} 
F^{\sin(\phi_h-\phi_S)}_{UT}(x,z,\bfPhp)  }
{2\pi^2\alpha^2\frac{1+(1-y)^2}{s x y^2}F_{UU}(x,z,\bfPhp)}\nonumber\\
=~~~~~&&\!\!\!\!\!\!\!\!\!\!\!\! \frac{2\pi^2\alpha^2\frac{1+(1-y)^2}{s x y^2} 
\sum_\nu e^2_\nu \int d^2p_\perp \{\frac{-\hat{\bf{P}}_{h\perp}.\bfp}{M}\} 
f^{\perp\nu}_{1T}(x,\bfp^2) D^{h/\nu}_1(z,\bfPh-z\bfp)  }
{2\pi^2\alpha^2\frac{1+(1-y)^2}{s x y^2}\sum_\nu e^2_\nu \int d^2p_\perp  f^\nu_{1}(x,\bfp^2) D^{h/\nu}_1(z,\bfPh-z\bfp) }.\label{Siv_Asy} 
\ee

\begin{figure}[htbp]
\begin{minipage}[c]{0.98\textwidth}
\includegraphics[scale=.75]{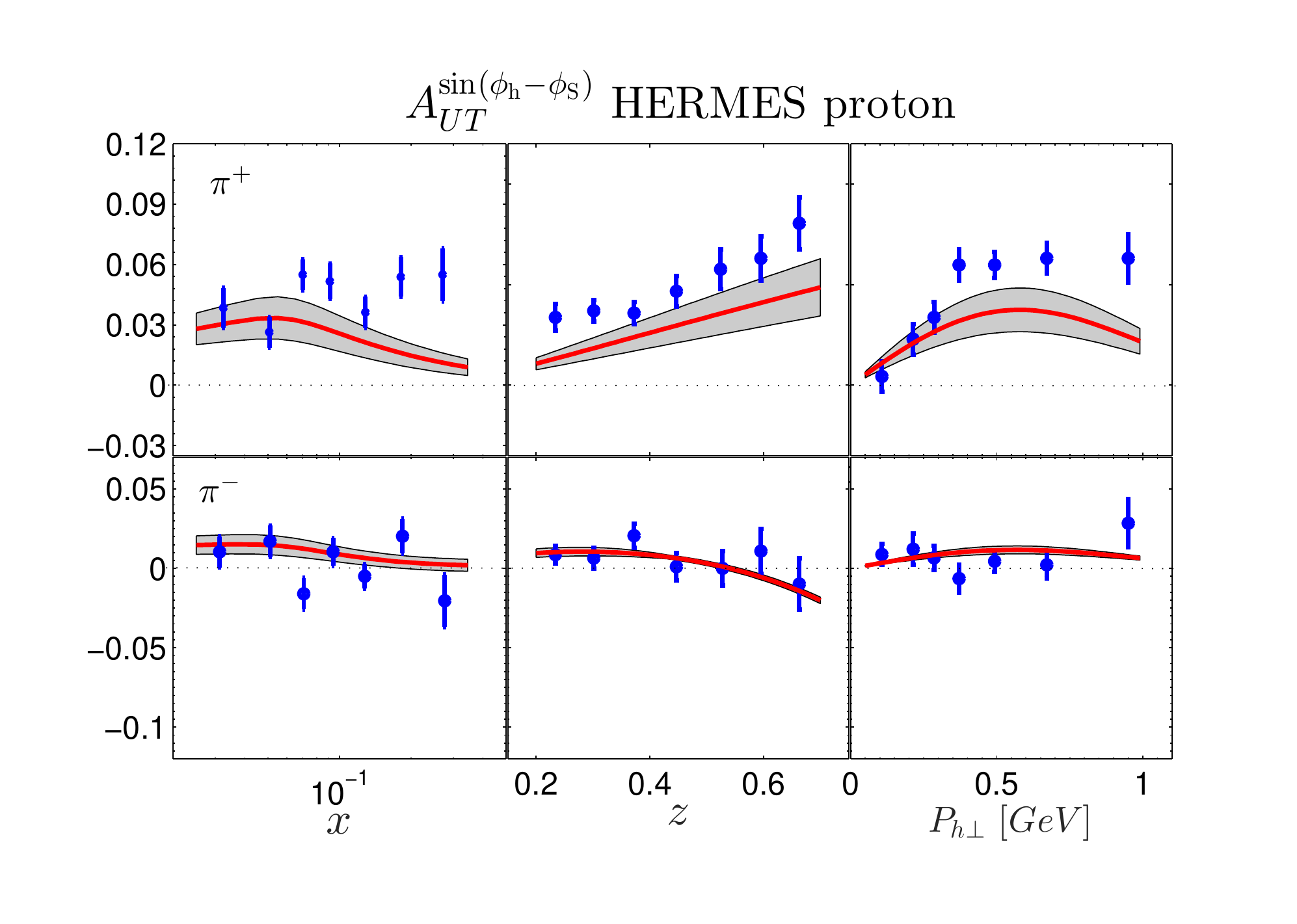} 
\end{minipage}
\caption{\label{fig_SivA_H} Model result of Sivers asymmetries, $A^{{\rm 
sin}(\phi_h-\phi_S)}_{UT}$, are shown by the continuous (red) lines for 
$\pi^+$(upper row) and $\pi^-$(lower row) channels and compared with the HERMES 
data\cite{Airapetian:2009ae}. $f^{\perp \nu}_{1T}(x,\bfp^2)$ are taken at 
initial scale and $f^\nu_1(x,\bfp^2)$ are evolved  to $\mu^2=2.5~GeV^2$ 
following the QCD evolution\cite{Aybat:2011zv}. The fragmentation function 
$D^{h/\nu}_1(z,\bfk)$ are taken as a phenomenological\cite{Kretzer:2001pz} input 
at $\mu^2=2.5~GeV^2$.}
\end{figure}

In this model, the explicit form of the Sivers functions is given in 
Eq.\ref{siv_TMD} and the unpolarized TMDs is given in \cite{Maji:2017bcz}. The 
model result for Sivers asymmetries are shown in Fig.\ref{fig_SivA_H} in the 
$\pi^+$ and $\pi^-$ channels and compared with the HERMES data 
\cite{Airapetian:2009ae} in the kinematical region 
\be 
0.023 < x < 0.4, \hspace{.6cm}
0.2 < z < 0.7, \hspace{.6cm}
0.31< y <0.95,\hspace{.6cm}
%\ee
{\rm and~~~} P_{h\perp}> 0.05~GeV.\ee 
To compare with the data, $f^{\perp \nu}_{1T}(x,\bfp^2)$ are taken at initial 
scale and $f^\nu_1(x,\bfp^2)$ are evolved to $\mu^2=2.5~GeV^2$ following the 
QCD evolution\cite{Aybat:2011zv}.  For another apporach of QCD evolution of 
 TMDs, see \cite{Echevarria:2014xaa,Echevarria:2012pw}.
 Though qualitatively our results agree with the HERMES data, but our 
predictions in $\pi^+$ channel is a bit smaller than the data whereas for 
$\pi^-$ channel, the model predictions are in good agreement with the data. 
This may be due to the fact that our model prediction for the Sivers function 
for $u$-quark is smaller than the phenomenological fit(see Fig. 
\ref{fig_siv_1GeV}).

\begin{figure}[htbp]
\begin{minipage}[c]{0.98\textwidth}
\includegraphics[scale=.75]{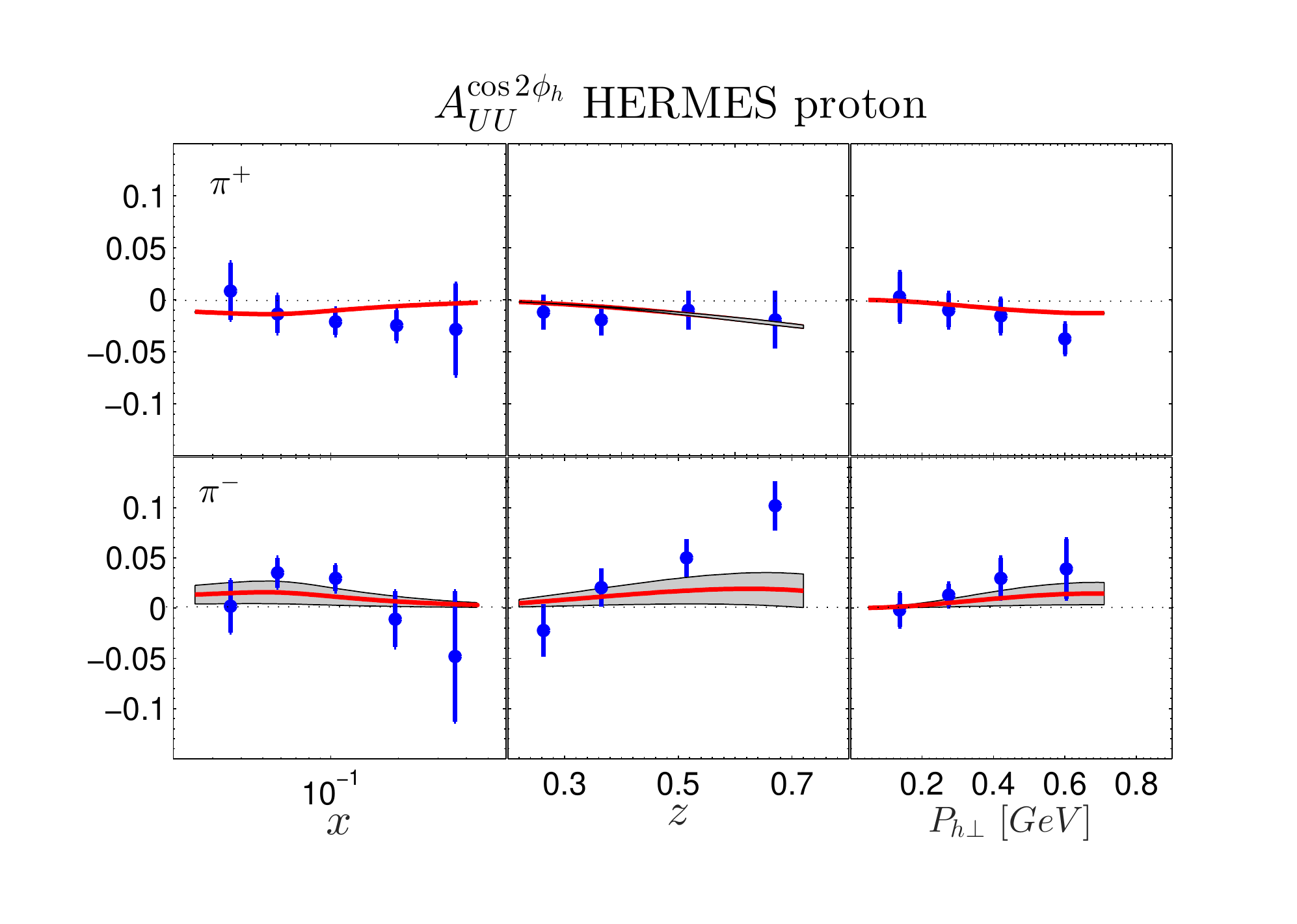} 
\end{minipage}
\caption{\label{fig_BMA_H} Model result of Boer-Mulders asymmetries, $A^{\cos 2\phi_h}_{UU}$. The continuous (red) lines represent the model prediction and the data are measured by HERMES collaboration\cite{Barone:2009hw,Giordano:2009hi}. $h^{\perp \nu}_{1}(x,\bfp^2)$ are taken at initial scale and $f^\nu_1(x,\bfp^2)$ are evolved at $\mu^2=2.5~GeV^2$ following the QCD evolution\cite{Aybat:2011zv}. The fragmentation function $H^{\perp\nu}_1(z,\bfk)$ are taken as a phenomenological\cite{Anselmino:2013vqa, Anselmino:2007fs} input at 
$\mu^2=2.5~GeV^2$.}
\end{figure}

\begin{figure}[htbp]
\begin{minipage}[c]{0.98\textwidth}
\includegraphics[scale=.75]{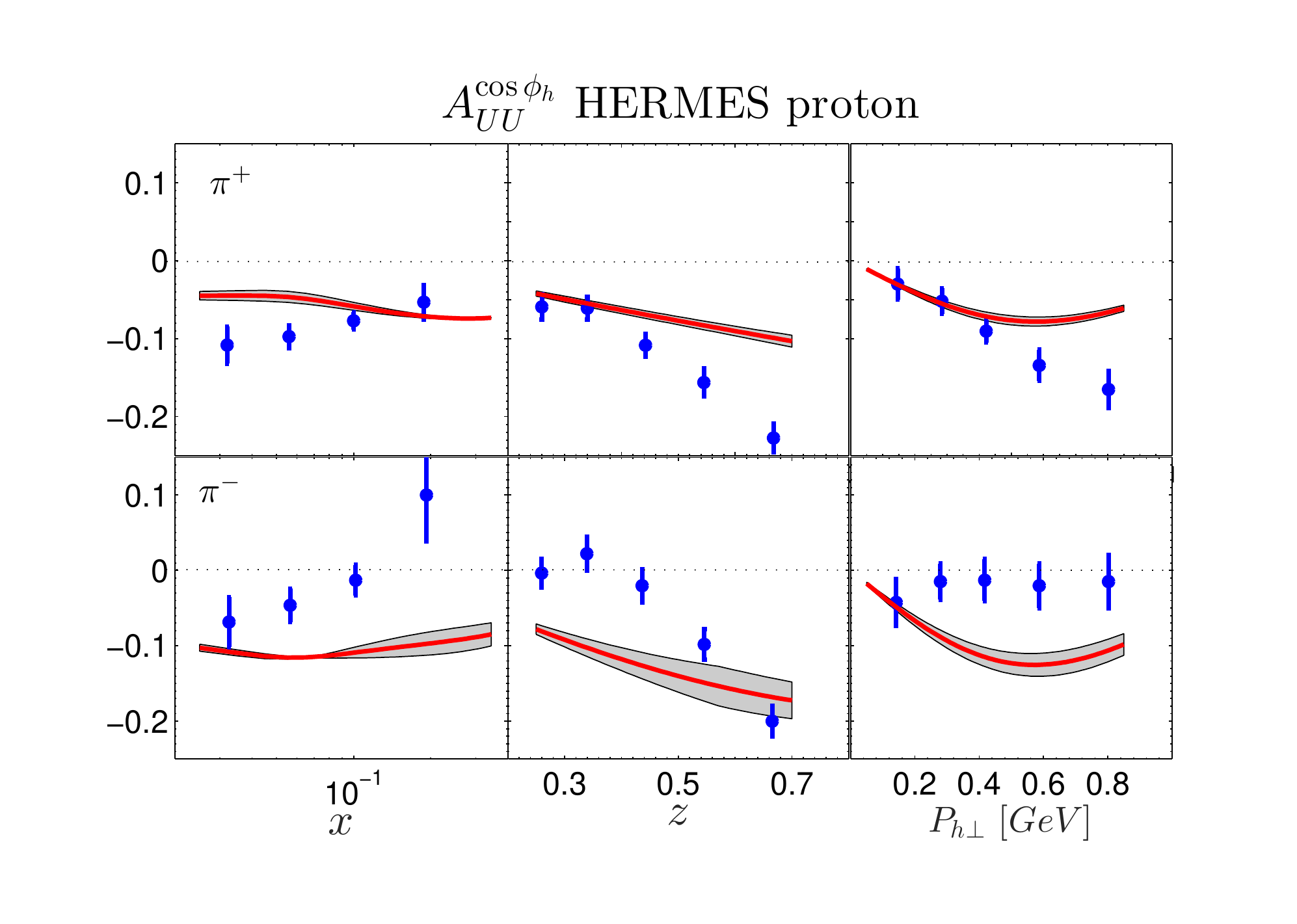} 
\end{minipage}
\caption{\label{fig_CahnA_H} Model results for  $A^{\cos \phi_h}_{UU}$ are 
shown by the continuous (red) lines and compared with HERMES 
data\cite{Airapetian:2012yg}. $h^{\perp \nu}_{1}(x,\bfp)$ are taken at initial 
scale and $f^\nu_1(x,\bfp)$ are evolved at $\mu^2=2.5~GeV^2$ following the QCD 
evolution\cite{Aybat:2011zv}. The fragmentation function 
$H^{\perp\nu}_1(z,\bfk)$ are taken as a phenomenological\cite{Anselmino:2013vqa, 
Anselmino:2007fs} input at $\mu^2=2.5~GeV^2$.}
\end{figure}

The Boer-Mulders asymmetry can be projected out replacing the weight factor in the Eq.\ref{Asy_def} by $\cos(2\phi_h)$ and written in terms of structure functions \cite{Anselmino:2011ch} as
\be 
A^{\cos (2\phi_h)}_{UU}&=& \frac{4\pi^2\alpha^2\frac{(1-y)}{s x y^2} F^{\cos 
2\phi_h}_{UU}(x,z,\bfPhp)  }
{2\pi^2\alpha^2\frac{1+(1-y)^2}{s x y^2}F_{UU}(x,z,\bfPhp)} \nonumber\\
&=& \frac{4\pi^2\alpha^2\frac{(1-y)}{s x y^2} \sum_\nu e^2_\nu \int d^2p_\perp \{\frac{(\bfPhp.\bfp)-2 z (\hat{\bf{P}}_{h\perp}.\bfp)^2 +z p^2_\perp }{z M_h M}\} h^{\perp\nu}_{1}(x,\bfp^2) H^{\perp\nu}_1(z,|\bfPh-z\bfp|)  }
{2\pi^2\alpha^2\frac{1+(1-y)^2}{s x y^2}\sum_\nu e^2_\nu \int d^2p_\perp  f^\nu_{1}(x,\bfp^2) D^{h/\nu}_1(z,|\bfPh-z\bfp|) }. \nonumber\\
\ee
The Boer-Mulders function in this model is given in Eq.\ref{BM_TMD}. We use the 
unpolarised fragmentation and the Collins  function 
$H^{\perp \nu}_1(z,\bfk)$ as  phenomenological inputs \cite{Anselmino:2013vqa, 
Kretzer:2001pz}.
\be 
D^{h/\nu}_1(z,\bfk)&=&D^{h/\nu}_1(z)\frac{e^{-\bfk^2/\avksq}}{\pi \avksq},\label{FF_D1}\\
 H^{\perp\nu}_1(z,\bfk)&=&(\frac{z M_h}{2 k_\perp}) 2\mathcal{N}^C_\nu(z) D^{h/\nu}_1(z)h(k_\perp)\frac{e^{-\bfk^2/\avksq}}{\pi \avksq}, \label{FF_H1}
\ee
with 
\be 
\mathcal{N}^C_\nu(z)&=&N^C_\nu z^{\rho_1} (1-z)^{\rho_2}\frac{(\rho_1+\rho_2)^{(\rho_1+\rho_2)}}{\rho^{\rho_1}_1\rho^{\rho_2}_2},\\
h(k_\perp)&=&\sqrt{2 e} \frac{k_\perp}{M_h}e^{-\bfk^2/M^2_h}.
\ee
Where $z=P_h^-/k^-$ is the energy fraction carried by the fragmenting quark of 
momentum $\textbf{k}$. The values of the parameters are listed in 
\cite{Anselmino:2013vqa} and  $D^{h/\nu}_1(z)$ is taken from the 
phenomenological extraction\cite{Kretzer:2001pz}.
Our model prediction to Boer-Mulders  asymmetry is shown in Fig.\ref{fig_BMA_H}. We compare our model result with the HERMES data\cite{Barone:2009hw,Giordano:2009hi} in the kinematical region
\be 
0.023 < x < 1.0, \hspace{.6cm}
0.2 < z < 1.0, \hspace{.6cm}
0.3 < y <0.85,\hspace{.6cm}
{\rm and~~~} P_{h\perp}> 0.05~GeV.\ee 
 The Boer-Mulders asymmetries agree with the HERMES data within error 
bars, except the plot with respect to $z$ in $\pi^-$ channel(Fig.\ref{fig_BMA_H}). The $A^{\cos(2\phi_h)}_{UU}$ asymmetry gets a 
twist-4 contribution due to Cahn 
effect\cite{Barone:2009hw} which is not included here.

Similarly, $\cos(\phi_h)$-weighted asymmetry also receives contribution 
from Cahn effect and Boer-Mulders function and is
defined \cite{Anselmino:2011ch} by
\be 
A^{\cos \phi_h}_{UU}&=& \frac{4\pi^2\alpha^2\frac{(2-y)\sqrt{(1-y)}}{s x y^2} 
F^{\cos \phi_h}_{UU}(x,z,\bfPhp)  }
{2\pi^2\alpha^2\frac{1+(1-y)^2}{s x y^2}F_{UU}(x,z,\bfPhp)} \nonumber\\
&=& \frac{4\pi^2\alpha^2\frac{(1-y)}{s x y^2} \big(-\frac{2}{Q}\big)\sum_\nu e^2_\nu \int d^2p_\perp [(\hat{\bf{P}}_{h\perp}.\bfp)f^\nu_1 D^{h/\nu}_1 +\frac{ p^2_\perp(P_{h\perp} - z \hat{\bf{P}}_{h\perp}.\bfp) }{z M_h M} h^{\perp\nu}_{1} H^{\perp\nu}_1]}{2\pi^2\alpha^2\frac{1+(1-y)^2}{s x y^2}\sum_\nu e^2_\nu \int d^2p_\perp  f^\nu_{1}(x,\bfp^2) D^{h/\nu}_1(z,|\bfPh-z\bfp|) }.
\ee
The model result for the asymmetry $A_{UU}^{\cos(\phi_h)}$ is shown in 
Fig.\ref{fig_CahnA_H} for $\pi^+$ and $\pi^-$ channels and compared with the 
HERMES data\cite{Airapetian:2012yg}. 
Recently, Boer-Mulders function and the 
Cahn effects have been extracted from the experimental data of 
$\cos 2\phi_h$ and $\cos \phi_h$ weighted asymmetries \cite{Christova:2017zxa}.

%%%%%%%%%%%%%%%%%%%%%%%
\section{Spin densities}
%%%%%%%%%%%%%%%%%%%%%%%
The spin density of unpolarised quarks with flavor $\nu$ in a transversely 
polarized proton is defined \cite{Bacchetta:2004jz} as
\be
f_{\nu/P^\uparrow}(x,\bfp)= f^\nu_1(x,\bfp^2) - \frac{\bf{S}.(\hat{\bf{P}} \times \bfp )}{M} f^{\perp \nu}_{1T}(x,\bfp^2). \label{SpinD_Siv}
\ee
\begin{figure}[htbp]
\begin{minipage}[c]{0.98\textwidth}
\includegraphics[width=7.5cm,clip]{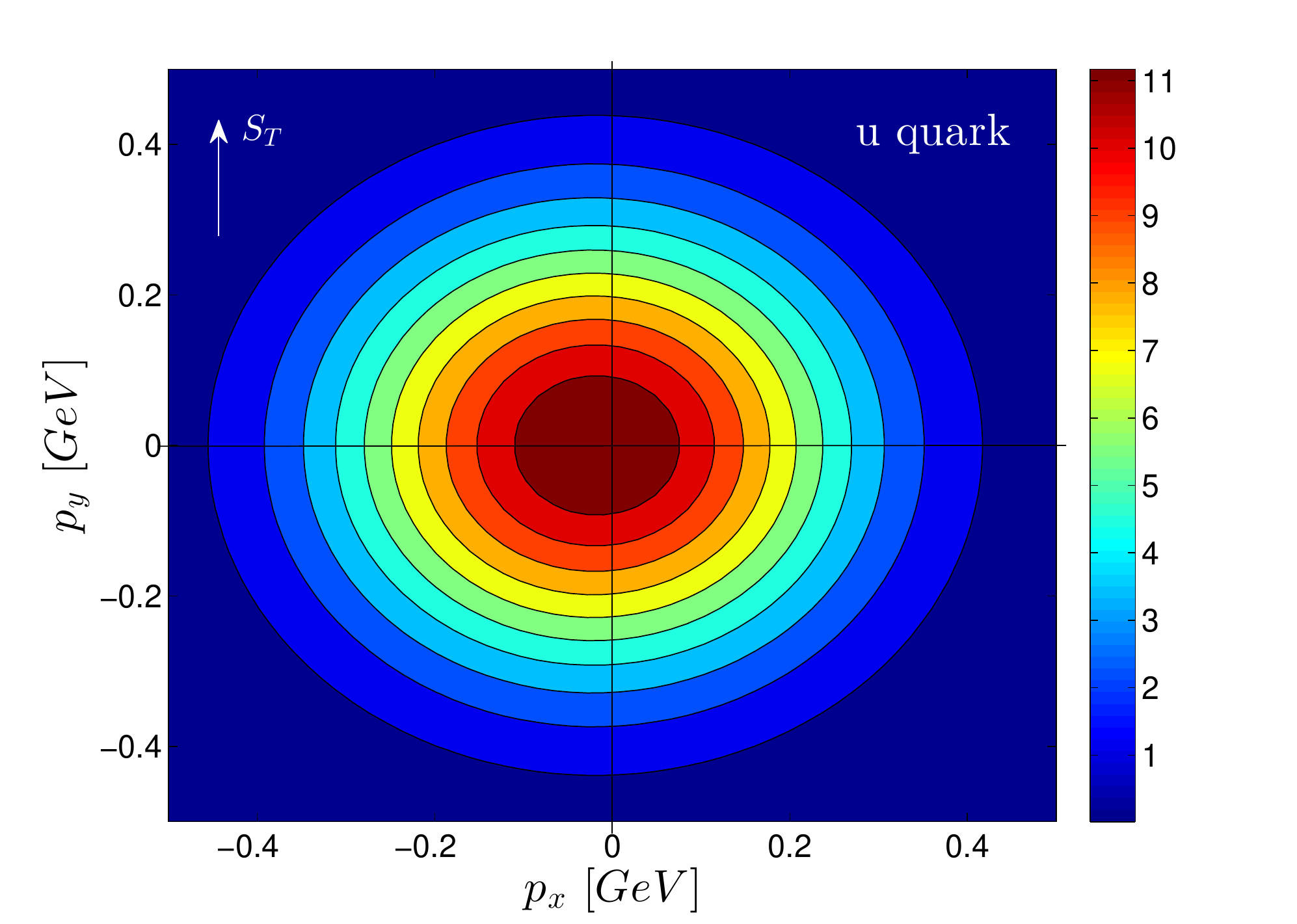} 
\includegraphics[width=7.5cm,clip]{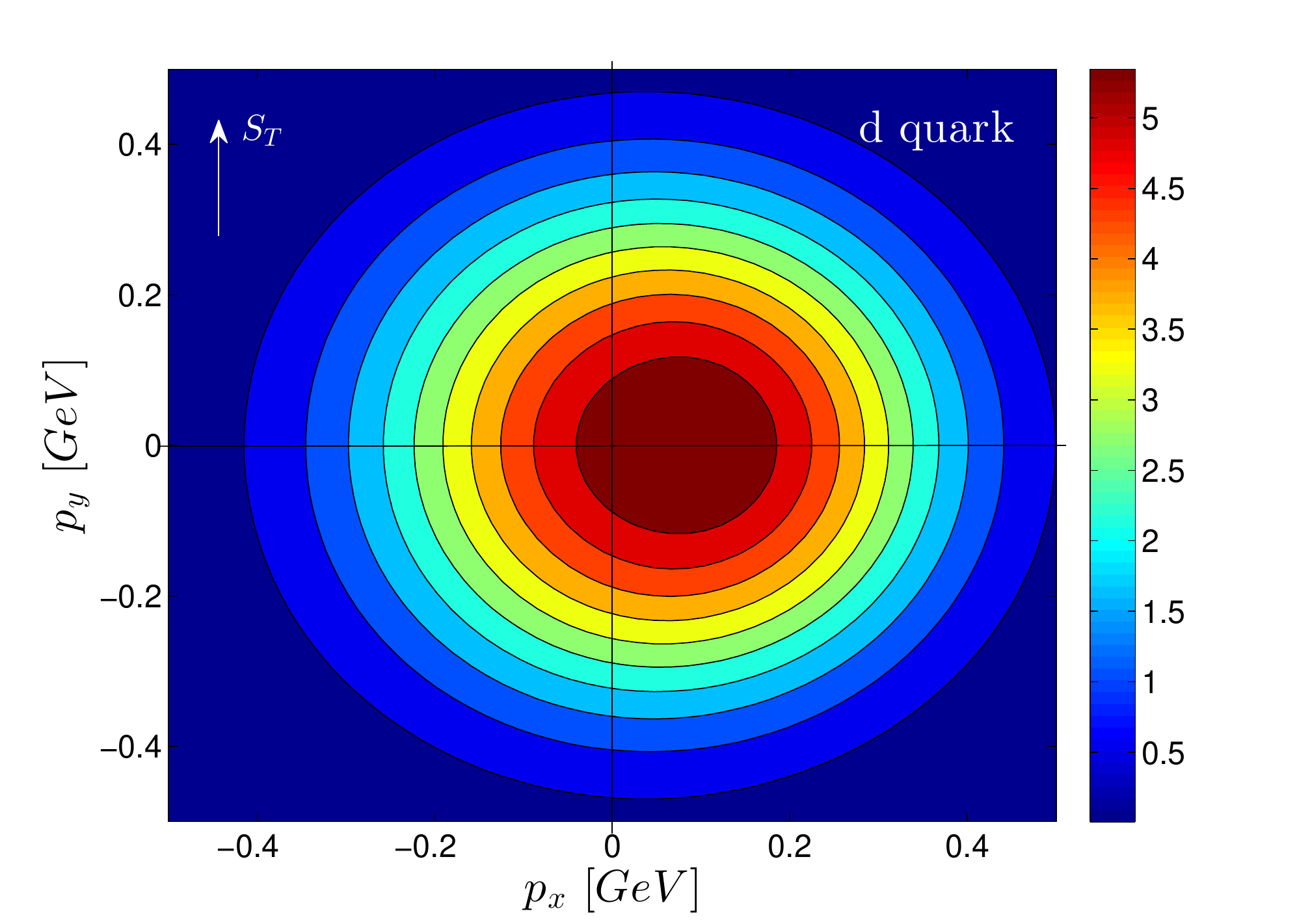}
\end{minipage}
\caption{\label{fig_SpinD_Siv} Spin density $f_{\nu/P^\uparrow}(x,\bfp)$ (Eq.\ref{SpinD_Siv}) are shown in transverse momentum plane for $u$ and $d$ quarks with $x=0.2$. The proton spin $\bf{S}$ is along the $y$-axis and the momentum of the proton $\bf{P}$ is along the $z$-direction.}  
\end{figure}

Here for SIDIS process, we take $\hat{\bf{P}}$ is along the direction of the 
momentum transfer $\hat{z}$-axis and the the spin of the proton $\bf{S}$ is 
along $\hat{\bf{y}}$. The spin density $f_{\nu/p^\uparrow}(x,\bfp)$ in the 
transverse momentum plane are shown in Fig.\ref{fig_SpinD_Siv} for both $u$ and 
$d$ quarks. Where the longitudinal momentum fraction $x=0.2$. The distributions 
are not symmetric, rather distorted towards left for $u$ quark and towards right 
for $d$ quark. This left-right distortions in the distribution is observed first 
time by D. Sivers \cite{Sivers:1989cc} and can be explained by the non-vanishing 
Sivers function $f^{\perp \nu}_{1T}(x,\bfp^2)$.  This is known as Sivers effect, 
where the quarks in a transversely polarized target have a transverse momentum 
asymmetry in the perpendicular direction to the nucleon spin $\bf{S}$. The left 
distortion is due to the negative distribution of Sivers functions for $u$ quark 
and the right distortion is due to the positive distribution of Sivers function 
for $d$ quark. Similar kind of distortions are observed in other model 
calculations \cite{Bacchetta:2008af} as well as in lattice QCD 
\cite{Gockeler:2006zu}.

\begin{figure}[htbp]
\begin{minipage}[c]{0.98\textwidth}
\includegraphics[width=7.5cm,clip]{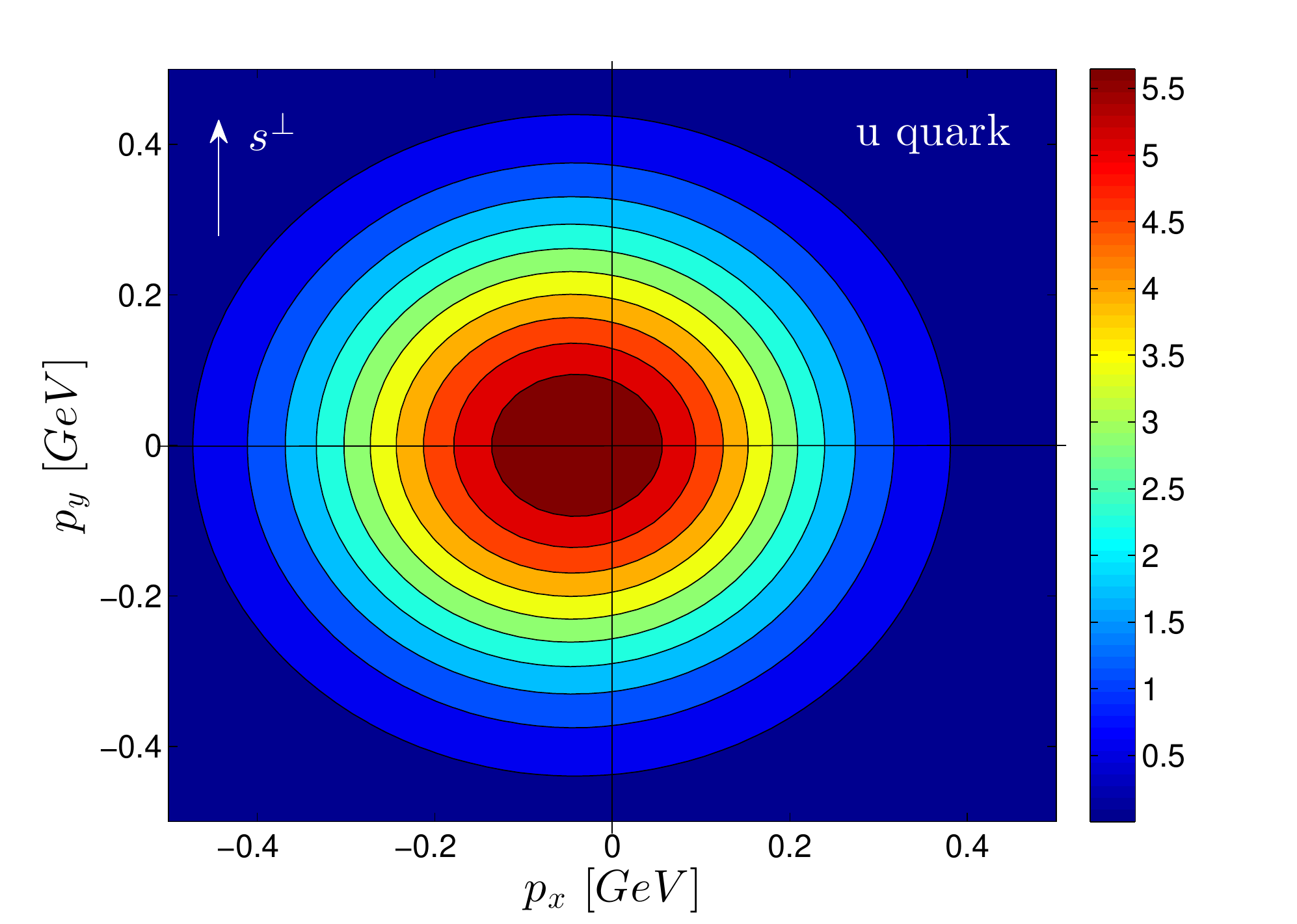} 
\includegraphics[width=7.5cm,clip]{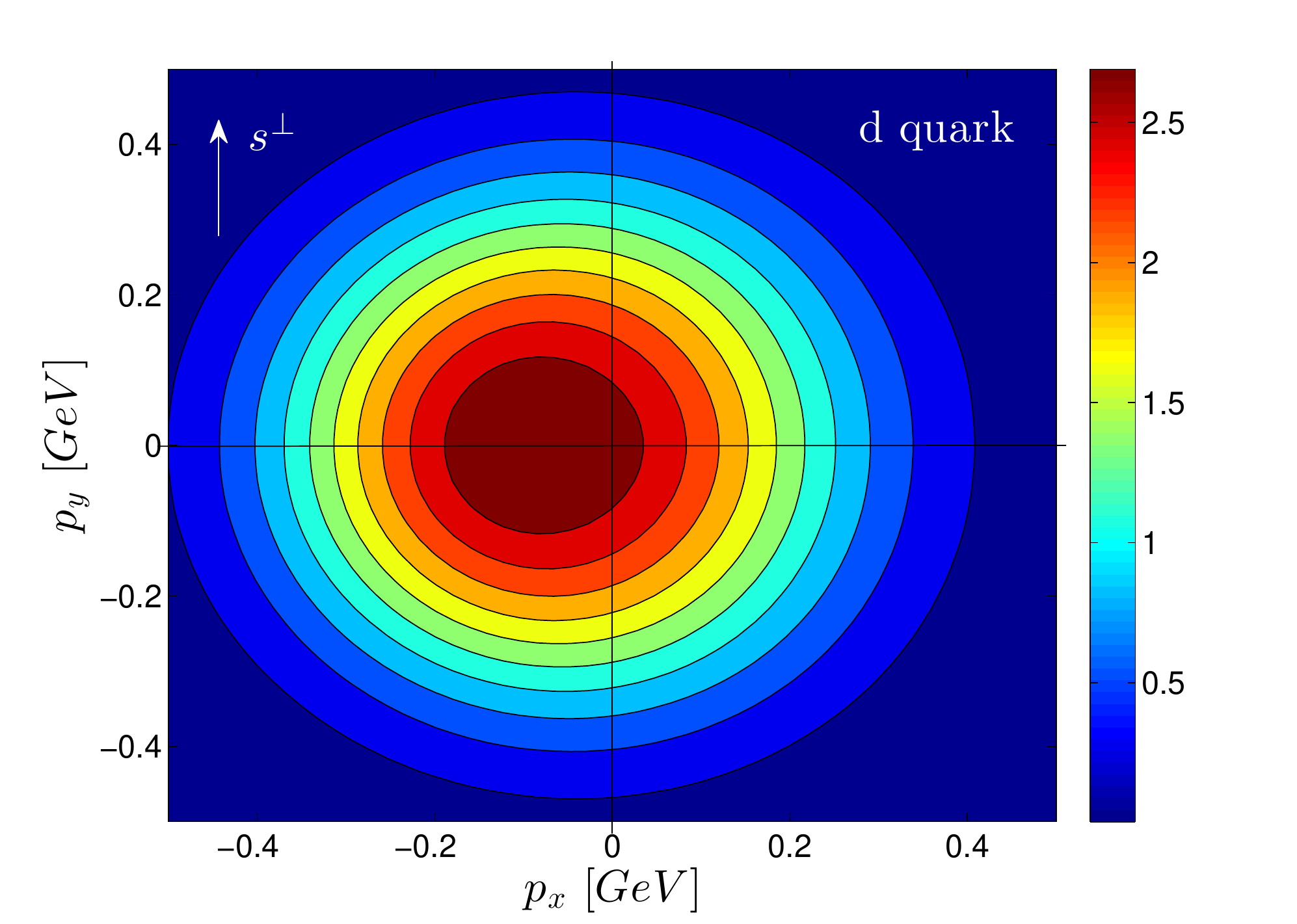}
\end{minipage}
\caption{\label{fig_SpinD_BM} Spin density $f_{\nu^\uparrow/P}(x,\bfp)$ (Eq.\ref{SpinD_BM}) are shown in transverse momentum plane for $u$ and $d$ quarks with $x=0.2$. The quark spin $\bf{s}$ is along the $y$-axis and the momentum of the proton $\bf{P}$ is along the $z$-direction.}
\end{figure}

Similarly, the spin density for transversely polarized quarks with flavor $\nu$ 
in an  unpolarized proton is defined \cite{Bacchetta:2004jz} as
\be
f_{\nu^\uparrow/P}(x,\bfp)= \frac{1}{2} [f^\nu_1(x,\bfp^2) - \frac{\bf{s}.(\hat{\bf{P}} \times \bfp )}{M} h^{\perp \nu}_{1}(x,\bfp^2)]. \label{SpinD_BM}
\ee
Where $\bf{s}$ is the spin of the interior quark. The spin density $f_{\nu^\uparrow/P}(x,\bfp)$ is shown in Fig.\ref{fig_SpinD_BM} for quark spin $\bf{s}$ along $\hat{\bf{y}}$ with $x=0.2$. Since Boer-Mulders functions are negative for both $u$ and $d$ quarks, we observed only a left-shift, unlike Sivers effect, as obtained in Fig.\ref{fig_SpinD_BM}. 

%The Boer-mulders effect suggest that transversely polarised quarks in a transversely polarised proton have a transverse momentum asymmetry along the perpendicular direction of the quark spin $\bf{s}$.

Sivers function is related with the anomalous magnetic moment and the orbital angular momentum of partons.
The Pauli form factor defined by the correlator with helicity-flip vector current, is written in terms of overlap representations as
\be 
-(q^1-iq^2) \frac{F^\nu_2(Q^2)}{2M} &=& \int^1_0 \frac{dx d^2 p_\perp}{16 \pi^3} \bigg[ C^2_S \sum_{\lambda_q}\sum_{\lambda_N \neq \lambda^\prime_N} \psi^{\lambda_N \dagger}_{\lambda_q}(x,\bfp)\psi^{\lambda^\prime_N}_{\lambda_q}(x,\bfp) \nonumber\\
&&\hspace{1.5cm} +  C^2_A \sum_{\lambda_q} \sum_{\lambda_D}\sum_{\lambda_N \neq \lambda^\prime_N} \psi^{\lambda_N \dagger}_{\lambda_q \lambda_D}(x,\bfp)\psi^{\lambda^\prime_N}_{\lambda_q \lambda_D}(x,\bfp)\bigg] \label{F2_def}\\
&=& \int^1_0 dx  \bigg(C^2_S N^{\nu 2}_S -C^2_A \frac{1}{3}N^{\nu 2}_0 \bigg) 2 T^\nu_3(x) (1-x)^3 e^{-Q^2 \frac{\ln(1/x)}{4 \kappa^2}}
\ee 
The anomalous magnetic moment $\kappa^\nu$ can be found from the Pauli form factor in the limit $Q^2=0$, $\kappa^\nu=F^\nu_2(0)$. Thus 
\be 
\kappa^\nu= \int^1_0 dx \kappa^\nu(x) = \int^1_0 dx  \bigg(C^2_S N^{\nu 2}_S -C^2_A \frac{1}{3}N^{\nu 2}_0 \bigg) 2 T^\nu_3(x) (1-x)^3.
\ee
A simple relation between integrated Sivers function(over $\bfp$) and anomalous magnetic moments is found as
\be 
f^{\perp\nu}_{1T}(x) = - C_F \alpha_s \mathcal{G}^\nu(x) \kappa^\nu(x) \label{Siv_k}
\ee
In this model, the relation can not be derived analytically, however numerical calculation gives the lensing function as
\be 
\mathcal{G}^\nu(x) \simeq \frac{1}{4 (1-x)}\bigg|_{\nu=u,d}
\ee
Similar type of lensing function is found in \cite{Lu:2006kt}. In the Ref.\cite{Bacchetta:2011gx}, $\mathcal{G}^\nu(x) \propto 1/(1-x)^\eta $ where $\eta$ is typically around 0.4  but $\eta$ can vary between 0.03 and 2. 

The total longitudinal angular momentum of parton $\nu$ is defined in terms of the moment of the GPDs as
\be 
J^\nu = \frac{1}{2} \int^1_0 dx x [H^\nu(x,0,0) + E^\nu(x,0,0)].
\ee
In the forward limit, moment of the $E$ and $H$ GPDs satisfy  
\be 
\int^1_0 dx H^\nu(x,0,0)&=& n^\nu = \int^1_0 dx d^2\bfp f^\nu_1(x,\bfp^2)\\
\int^1_0 dx E^\nu(x,0,0)&=& \kappa^\nu
\ee
Where $n^u=2$ and $n^d=1$ for proton. From iso-spin symmetry flavored anomalous magnetic moments are $\kappa^u=1.673$ and $\kappa^d=-2.033$. GPDs are discussed in this model \cite{Maji:2017ill}. We define $\kappa^\nu=\int dx \kappa^\nu(x)$ and $ \kappa^\nu(x)= E^\nu(x,0,0)$.  
Therefore, the Eq.\ref{Siv_k} is modified as
\be 
f^{\perp\nu}_{1T}(x) \simeq - C_F \alpha_s \frac{1}{4 (1-x)} E^\nu(x,0,0)\label{Siv_E}
\ee
Thus the longitudinal angular momentum can be calculated from the moment of 
Sivers function and unpolarised TMDs as
\be 
J^\nu = \frac{1}{2} \int^1_0 dx x [f_1^\nu(x) - \frac{4(1-x)}{C_F \alpha_s} f^{\perp \nu}_{1T}(x)].
\ee
In this model, we obtain 
\be 
J^u = 0.9559 ~~ {\rm and} ~~ J^d = -0.5791.
\ee
Total contribution to the nucleon spin from $u$ and $d$ quarks is $0.3768$ 
at the initial scale $\mu_0=0.8~ GeV$. Cloudy bag model \cite{Aidala:2012mv} 
and lattice 
calculations predict total angular momentum contribution about $0.24$ at a scale 
of $\mu^2=4~ GeV^2$.
%i.e., almost 75\% of the nucleon spin, at the initial scale.

%%%%%%%%%%%%%%%%%%%%%%%%%
\section{Conclusions}
%%%%%%%%%%%%%%%%%%%%%%%

We have presented the results for $T$-odd TMDs namely, the Sivers and 
Boer-Mulders functions in a light-front quark-diquark model of the proton and 
the spin asymmetries in SIDIS associated with these functions. It is well known 
that the  final state interaction is responsible to produce the required complex 
phase in the amplitude which gives rise to the Sivers asymmetries. Though the 
proton wave function in principle cannot describe FSI (as the participating 
quark comes out of the proton state), as Hwang\cite{Hwang:2010dd} proposed, we have 
modelled 
the light-front wave functions to incorporate the effects of the FSI. This is done 
by extending the wave functions in the quark-diquark model to have complex phases 
consistent with the SIDIS amplitudes.  The complex phases in the light-front 
wave functions produce the Sivers and Boer-Mulders functions. Both Sivers and 
Boer Mulders functions and their moments are evaluated in this model and 
compared with other model and phenomenological fits. The Sivers asymmetry 
$A_{UT}^{\sin(\phi_h-\phi_S)}$ for $\pi^+$ channel is found to be a bit smaller 
than the experimental data; better agreements are observed for Boer-Mulders 
asymmetry $A_{UU}^{\cos(2\phi_h)}$ for both $\pi^+$ and $\pi^-$ channels. 
Sivers and Boer-Mulders functions help us to understand the spin structure of 
the proton at the  parton level. Due to Sivers effect the spin density of 
an unpolarised quark in a transversely polarized proton is found to be 
asymmetric in the perpendicular direction to the nuclear spin. The distortions 
due to Sivers effect in our model for both  $u$ and $d$ quarks are consistent 
with the results found in other models and lattice QCD. Since the 
Sivers function is negative for $u$ and positive for $d$ quark, the 
distortion for $u$ quark  is in opposite direction of the $d$ quark. Similarly, 
Boer-Mulders function produces the distortion in the spin density of a 
transversely polarized quark in a transversely polarized proton. Since 
Boer-Mulders function has the same sign for both $u$ and $d$ quarks, the 
distortions in the spin densities are also in the same direction. Sivers 
function integrated over the transverse momentum is related to the anomalous 
magnetic moment through the lensing function. Our model predicts that the 
lensing function should go as $(1-x)^{-1}$.
%The model also predicts that the longitudinal angular momentum of 
%the quarks contributes around $0.38$ to the proton spin.

%%

\end{document}